\RequirePackage{lineno}  
\documentclass[a4paper,11pt]{article}
\usepackage{jheppub} 
\usepackage{atlasphysics}
\usepackage{cancel}   
\usepackage{float}
\usepackage{morefloats}
\usepackage{verbatim}
\usepackage{multirow}
\usepackage{harpoon}

\usepackage{preprintcover}  
\PreprintCoverPaperTitle{\boldmath Measurement of the distributions of event-by-event flow harmonics in lead--lead collisions at $\sqrt{s_{_{NN}}}=2.76$ TeV with the ATLAS detector at the LHC}

\PreprintIdNumber{CERN-PH-EP-2013-044}  

\PreprintCoverAbstract{The distributions of event-by-event harmonic flow coefficients $v_n$ for $n=$2--4 are measured in $\sqrt{s_{NN}}=2.76$ TeV Pb+Pb collisions using the ATLAS detector at the LHC. The measurements are performed using charged particles with transverse momentum $\pT> 0.5$~GeV and in the pseudorapidity range $|\eta|<2.5$ in a dataset of approximately 7~$\mu\mathrm{b}^{-1}$ recorded in 2010. The shapes of the $v_n$ distributions suggest that the associated flow vectors are described by a two-dimensional Gaussian function in central collisions for $v_2$ and over most of the measured centrality range for $v_3$ and $v_4$. Significant deviations from this function are observed for $v_2$ in mid-central and peripheral collisions, and a small deviation is observed for $v_3$ in mid-central collisions. In order to be sensitive to these deviations, it is shown that the commonly used multi-particle cumulants, involving four particles or more, need to be measured with a precision better than a few percent. The $v_n$ distributions are also measured independently for charged particles with $0.5<\pT<1$~GeV and $\pT>1$~GeV. When these distributions are rescaled to the same mean values, the adjusted shapes are found to be nearly the same for these two $\pT$ ranges. The $v_n$ distributions are compared with the eccentricity distributions from two models for the initial collision geometry: a Glauber model and a model that includes corrections to the initial geometry due to gluon saturation effects. Both models fail to describe the experimental data consistently over most of the measured centrality range.}
\PreprintJournalName{JHEP}  

\title{Measurement of the distributions of event-by-event flow harmonics in lead--lead collisions at $\sqrt{s_{_{NN}}}=2.76$ TeV with the ATLAS detector at the LHC}
\author{The ATLAS Collaboration}

\date{\today}

\abstract{
The distributions of event-by-event harmonic flow coefficients $v_n$ for $n=$2--4 are measured in $\sqrt{s_{NN}}=2.76$ TeV Pb+Pb collisions using the ATLAS detector at the LHC. The measurements are performed using charged particles with transverse momentum $\pT> 0.5$~GeV and in the pseudorapidity range $|\eta|<2.5$ in a dataset of approximately 7~$\mu\mathrm{b}^{-1}$ recorded in 2010. The shapes of the $v_n$ distributions suggest that the associated flow vectors are described by a two-dimensional Gaussian function in central collisions for $v_2$ and over most of the measured centrality range for $v_3$ and $v_4$. Significant deviations from this function are observed for $v_2$ in mid-central and peripheral collisions, and a small deviation is observed for $v_3$ in mid-central collisions. In order to be sensitive to these deviations, it is shown that the commonly used multi-particle cumulants, involving four particles or more, need to be measured with a precision better than a few percent. The $v_n$ distributions are also measured independently for charged particles with $0.5<\pT<1$~GeV and $\pT>1$~GeV. When these distributions are rescaled to the same mean values, the adjusted shapes are found to be nearly the same for these two $\pT$ ranges. The $v_n$ distributions are compared with the eccentricity distributions from two models for the initial collision geometry: a Glauber model and a model that includes corrections to the initial geometry due to gluon saturation effects. Both models fail to describe the experimental data consistently over most of the measured centrality range. }
\begin{document}
\maketitle
\section{Introduction}
Heavy ion collisions at the Relativistic Heavy Ion Collider (RHIC) and the Large Hadron Collider (LHC) create hot, dense matter that is thought to be composed of strongly interacting quarks and gluons. A useful tool to study the properties of this matter is the azimuthal anisotropy of particle emission in the transverse plane~\cite{Ollitrault:1992bk,Poskanzer:1998yz}. This anisotropy has been interpreted as a result of pressure-driven anisotropic expansion (referred to as ``flow'') of the created matter, and is described by a Fourier expansion of the particle distribution in azimuthal angle $\phi$, around the beam direction:
\begin{equation}
\label{eq:flow}
\frac{{\rm d}N}{{\rm d}\phi}\propto1+2\sum_{n=1}^{\infty}v_{n}\cos n(\phi-\Phi_{n})\;,
\end{equation}
where $v_n$ and $\Phi_n$ represent the magnitude and phase of the $n^{\mathrm{th}}$-order anisotropy of a given event in the momentum space. These quantities can also be conveniently represented by the per-particle ``flow vector''~\cite{Poskanzer:1998yz}: \mbox{$\overrightharp{v}_n=(v_n\cos n\Phi_{n},v_n\sin n\Phi_{n})$}. The angles $\Phi_n$ are commonly referred to as the event plane (EP) angles.

In typical non-central~\cite{Poskanzer:1998yz} heavy ion collisions, the large and dominating $v_2$ coefficient is associated mainly with the ``elliptic'' shape of the nuclear overlap. However, $v_2$ in central (head-on) collisions and the other $v_n$ coefficients in general are related to various shape components of the initial state arising from fluctuations of the nucleon positions in the overlap region~\cite{Alver:2010gr}. The amplitudes of these shape components, characterized by eccentricities $\epsilon_n$, can be estimated via a simple Glauber model from the transverse positions $(r,\phi)$ of the participating nucleons relative to their centre of mass~\cite{Qin:2010pf,Teaney:2010vd}:
\begin{eqnarray}
\label{eq:ena}
\epsilon_n = \frac{\sqrt{\langle r^n\cos n\phi\rangle^2+\langle r^n\sin n\phi\rangle^2}}{\langle r^n\rangle}\;.
\end{eqnarray}
The large pressure gradients and ensuing hydrodynamic evolution can convert these shape components into $v_n$ coefficients in momentum space.  Calculations based on viscous hydrodynamics suggest that $v_n$ scales nearly linearly with $\epsilon_n$, for $n<4$~\cite{Qiu:2011iv}. The proportionality constant is found to be sensitive to properties of the matter such as the equation of state and the ratio of shear viscosity to entropy density~\cite{Voloshin:2008dg,Teaney:2009qa}. In particular, the proportionality constant is predicted to decrease quickly with increasing shear viscosity~\cite{Alver:2010dn}. Hence detailed measurements of $v_n$ coefficients and comparisons with $\epsilon_n$ may shed light on the collision geometry of the initial state and transport properties of the created matter~\cite{Gale:2012rq,Niemi:2012aj}.

Significant $v_n$ coefficients have been observed for $n\leq6$ at RHIC and the LHC~\cite{Adare:2011tg,star:2013wf,Aamodt:2011by,ATLAS:2011ah,Aad:2012bu,Chatrchyan:2012wg,Chatrchyan:2012ta}. These observations are consistent with small values for the ratio of shear viscosity to entropy density, and the existence of sizable fluctuations in the initial state. Most of these measurements estimate $v_n$ from the distribution of particles relative to the event plane, accumulated over many events. This event-averaged $v_n$ mainly reflects the hydrodynamic response of the created matter to the average collision geometry in the initial state. More information, however, can be obtained by measuring $\overrightharp{v}_n$ or $v_n$ on an event-by-event (EbyE) basis. 

Some properties of the $v_n$ distributions, such as the mean $\langle v_n\rangle$, the standard deviation (hereafter referred to as ``width'') $\sigma_{v_n}$, the relative fluctuation $\sigma_{v_n}/\langle v_n\rangle$, and the root-mean-square $\sqrt{\langle v_n^2\rangle}\equiv\sqrt{\langle v_n\rangle^2+\sigma_{v_n}^2}$, were previously estimated from a Monte Carlo template fit method~\cite{Alver:2007qw}, or two- and four-particle cumulant methods~\cite{Borghini:2000sa,Agakishiev:2011eq,Abelev:2012di}. The value of $\sigma_{v_2}/\langle v_2\rangle$ was measured to be 0.3--0.7 in different centrality bins. However, these methods are reliable only for $\sigma_{v_n}\muchless\langle v_n\rangle$, and are subject to significant systematic uncertainties. In contrast, $\langle v_n\rangle$, $\sigma_{v_n}$ and higher-order moments can be calculated directly from the full $v_n$ distributions. 

The EbyE distributions of $\overrightharp{v}_n$ or $v_n$ also provide direct insight into the fluctuations in the initial geometry~\cite{Voloshin:2007pc}. If fluctuations of $\overrightharp{v}_n$ relative to the underlying flow vector associated with the average geometry, $\overrightharp{v}_n^{\mathrm{RP}}$, in the reaction plane~\footnote{The reaction plane is defined by the impact parameter vector and the beam axis.} (RP)~\cite{Bhalerao:2006tp,Voloshin:2007pc} are described by a two-dimensional (2D) Gaussian function in the transverse plane, then the probability density of $\overrightharp{v}_n$ can be expressed as:
\begin{eqnarray}
\label{eq:fluc1}
p(\overrightharp{v}_n) &=& \frac{1}{2\pi\delta^2_{_{v_n}}} e^{-\left(\overrightharp{v}_n-\overrightharp{v}_n^{\mathrm{\;RP}}\right)^2 \big{/}\left(2\delta^2_{_{v_n}}\right)}\;.
\end{eqnarray}
Model calculations show that this approximation works well for central and mid-central collisions~\cite{Voloshin:2007pc,Alver:2008zza}. Integration of this 2D Gaussian over the azimuthal angle gives the one-dimensional (1D) probability density of $v_n=|\protect\overrightharp{v}_n|$ in the form of the Bessel--Gaussian function~\cite{Voloshin:1994mz,Voloshin:2008dg}:
\begin{eqnarray}
\label{eq:fluc2}
p(v_n) =\frac{v_n}{\delta_{_{v_n}}^2}e^{-\frac{(v_n)^2+(v_n^{\mathrm{RP}})^2}{2\delta_{_{v_n}}^2}} I_0\left(\frac{v_n^{\mathrm{RP}}v_n}{\delta_{_{v_n}}^2}\right)\;,
\end{eqnarray}
where $I_0$ is the modified Bessel function of the first kind. Additional smearing to eq.~(\ref{eq:fluc1}) also arises from effects of the finite number of particles produced in the collision. If it is Gaussian, this smearing is expected to increase the observed $\delta_{_{v_n}}$ value, but the value of $v_n^{\mathrm{RP}}$ should be stable.

The parameters $v_n^{\mathrm{RP}}$ and $\delta_{_{v_n}}$ in eq.~(\ref{eq:fluc2}) are related to $\langle v_n\rangle$ and $\sigma_{v_n}$, and can be estimated directly from a fit of the measured $p(v_n)$ distribution with eq.~(\ref{eq:fluc2}). For small fluctuations $\delta_{_{v_n}}\muchless v_n^{\mathrm{RP}}$~\cite{Voloshin:2007pc}:
\begin{eqnarray}
\label{eq:small}
 \delta_{_{v_n}}\approx \sigma_{v_n},\;\;\; \left(v_n^{\mathrm{RP}}\right)^2\approx\langle v_n\rangle^2-\delta^2_{_{v_n}}\;.
\end{eqnarray}
For large fluctuations $\delta_{_{v_n}}\gg v_n^{\mathrm{RP}}$ (e.g. in central collisions), eqs.~(\ref{eq:fluc1}) and (\ref{eq:fluc2}) can be approximated by:
\begin{eqnarray}
\label{eq:fluc2b}
p(\overrightharp{v}_n) = \frac{1}{2\pi\delta^2_{_{v_n}}} e^{-\overrightharp{v}^{\mathrm{\;2}}_n/\left(2\delta^2_{_{v_n}}\right)},\; p(v_n) = \frac{v_n}{\delta^2_{_{v_n}}} e^{-v^{\mathrm{\;2}}_n/\left(2\delta^2_{_{v_n}}\right)},
\end{eqnarray}
which is equivalent to the ``fluctuation-only'' scenario, i.e. $v_n^{\mathrm{RP}}=0$. In this case, both the mean and the width are controlled by $\delta_{_{v_n}}$~\cite{Broniowski:2007ft}:
\begin{eqnarray}
\label{eq:fluc3}
\langle v_n\rangle =\sqrt{\frac{\pi}{2}}\;\delta_{_{v_n}},\;\sigma_{v_n} = \sqrt{2-\frac{\pi}{2}}\;\delta_{_{v_n}},
\end{eqnarray}
and hence:
\begin{eqnarray}
\label{eq:large}
\frac{\sigma_{v_n}}{\langle v_n\rangle} = \sqrt{\frac{4}{\pi}-1}=0.523,\; \sqrt{\langle v_n^2\rangle}=\frac{2}{\sqrt{\pi}}\langle v_n\rangle=1.13\langle v_n\rangle\;.
\end{eqnarray}
In the intermediate case, $\delta_{_{v_n}}\approx v_n^{\mathrm{RP}}$, a more general approximation to eq.~(\ref{eq:fluc2}) can be used via a Taylor expansion of the Bessel function, $I_0(x)=e^{x^2/4}\left[1-x^4/64+O(x^6)\right]$:
\begin{eqnarray}
\label{eq:fluc2c}
p(v_n) &\approx&\frac{v_n}{\delta^{\prime 2}_{_{v_n}}} e^{-v^{\mathrm{\;2}}_n/\left(2\delta^{\prime 2}_{_{v_n}}\right)} {\textstyle \left[1-\left(\frac{v_n^{\mathrm{RP}}v_n}{\delta_{_{v_n}}^2}\right)^4/64\right]},\\\label{eq:fluc2d}
\delta^{\prime 2}_{_{v_n}} &=& \delta^{2}_{_{v_n}} {\textstyle \left(1-\frac{(v_n^{\mathrm{RP}})^2}{2\delta^{2}_{_{v_n}}}\right)}^{-1}\approx \delta^{2}_{_{v_n}}+ (v_n^{\mathrm{RP}})^2/2\;.
\end{eqnarray}
Defining $\alpha\equiv \delta_{_{v_n}}/v_n^{\mathrm{RP}}$, eqs.~(\ref{eq:fluc2c}) and (\ref{eq:fluc2d}) imply that for $v_n\muchless 2\sqrt{2}\delta_{_{v_n}} \alpha$, the shape of the distribution is very close to that of eq.~\ref{eq:fluc2b}, except for a redefinition of the width. For example, the deviation from the fluctuation-only scenario is expected to be less than 10\% over the range $v_n< 1.6\delta_{_{v_n}} \alpha$. Hence the reliable extraction of $v_n^{\mathrm{RP}}$ requires precise determination of the tails of the $v_n$ distributions, especially when $v_n^{\mathrm{RP}}$ is smaller than $\delta_{_{v_n}}$. This is especially important for the study of the $v_3$ and $v_4$ distributions, which are expected to be dominated by $\delta_{_{v_n}}$.

Each quantity mentioned above, $\langle v_n\rangle$, $\sigma_{v_n}$, $\sqrt{\langle v_n^2\rangle}$, $\sigma_{v_n}/\langle v_n\rangle$, $v_n^{\mathrm{RP}}$ and $\delta_{_{v_n}}$, has been the subject of extensive studies both experimentally~\cite{Alver:2007qw,Abelev:2012di,Alver:2008zza} and in theoretical models~\cite{Bhalerao:2006tp,Voloshin:2007pc,Ollitrault:2009ie}. Experimental measurement of the EbyE $v_n$ distributions can elucidate the relations between these quantities,  as well as clarify the connections between various experimental methods. In particular, previous measurements based on multi-particle cumulant methods suggest that the $v_2$ distributions are consistent with the Bessel--Gaussian function~\cite{Adams:2004bi,Bilandzic:2011ww}. However, this consistency is inferred indirectly from agreement among four-, six- and eight-particle cumulants: the measurement of the EbyE $v_n$ distributions can directly quantify whether the Bessel--Gaussian function is the correct description of the data.

This paper presents the measurement of the EbyE distribution of $v_2$, $v_3$ and $v_4$ over a broad range of centrality in lead--lead (Pb+Pb) collisions at $\sqrt{s_{NN}}=2.76$ TeV with the ATLAS detector at the LHC. The observed $v_n$ distributions are measured using charged particles in the pseudorapidity range $|\eta|<2.5$ and the transverse momentum range $\pT>0.5$~GeV, which are then unfolded to estimate the true $v_n$ distributions. The key issue in the unfolding is to construct a response function via a data-driven method, which maps the true $v_n$ distribution to the observed $v_n$ distribution. This response function corrects mainly for the smearing due to the effect of finite charged particle multiplicity in an event, but it also suppresses possible non-flow effects from short-range correlations~\cite{Jia:2013tja}, such as resonance decays, Bose--Einstein correlations and jets~\cite{Voloshin:2008dg}.

The paper is organized as follows. Sections~\ref{sec:det} and \ref{sec:sel} give a brief overview of the ATLAS detector, trigger, and selection criteria for events and tracks. Section~\ref{sec:m1} discusses the details of the single-particle method and the two-particle correlation method used to obtain the observed $v_n$ values, the Bayesian unfolding method used to estimate the true distributions of $v_n$, and the performance of the unfolding procedure and systematic uncertainties of the measurement. The results are presented in section~\ref{sec:re}, and a summary is given in section~\ref{sec:con}.

\section{The ATLAS detector and trigger}
\label{sec:det}
The ATLAS detector~\cite{Aad:2008zzm} provides nearly full solid-angle coverage around the collision point with tracking detectors, calorimeters and muon chambers, which are well suited for measurements of azimuthal anisotropies over a large pseudorapidity range\footnote{ATLAS uses a right-handed coordinate system with its origin at the nominal interaction point (IP) in the centre of the detector and the $z$-axis along the beam pipe. The $x$-axis points from the IP to the centre of the LHC ring, and the $y$-axis points upward. Cylindrical coordinates $(r,\phi)$ are used in the transverse plane, $\phi$ being the azimuthal angle around the beam pipe. The pseudorapidity is defined in terms of the polar angle $\theta$ as $\eta=-\ln\tan(\theta/2)$.}. This analysis uses primarily two subsystems: the inner detector (ID) and the forward calorimeter (FCal). The ID is immersed in the 2~T axial field of a superconducting solenoid magnet, and measures the trajectories of charged particles in the pseudorapidity range $|\eta|<2.5$ and over the full azimuthal range. A charged particle passing through the ID traverses typically three modules of the silicon pixel detector (Pixel), four double-sided silicon strip modules of the semiconductor tracker (SCT) and, for $|\eta|<2$, a transition radiation tracker composed of straw tubes.  The FCal covers the range $3.1 < |\eta| < 4.9$ and is composed of symmetric modules at positive and negative $\eta$. The FCal modules are composed of either tungsten or copper absorbers with liquid argon as the active medium, which together provide ten interaction lengths of material. In heavy ion collisions, the FCal is used mainly to measure the event centrality and event plane angles~\cite{ATLAS:2011ah,Aad:2012bu}.

The minimum-bias Level-1 trigger used for this analysis requires signals in two zero-degree calorimeters (ZDC) or either of the two minimum-bias trigger scintillator (MBTS) counters. The ZDCs are positioned at 140~m from the collision point, detecting neutrons and photons with $|\eta|>8.3$, and the MBTS covers $2.1<|\eta|<3.9$ on each side of the nominal interaction point. The ZDC Level-1 trigger thresholds on each side are set below the peak corresponding to a single neutron. A Level-2 timing requirement based on signals from each side of the MBTS is imposed to remove beam-induced backgrounds.

\section{Event and track selections}
\label{sec:sel}
This paper is based on approximately 7~$\mu\mathrm{b}^{-1}$ of Pb+Pb collisions collected in 2010 at the LHC with a nucleon--nucleon centre-of-mass energy $\sqrt{s_{_{NN}}}=2.76$~TeV. An offline event selection requires a time difference $|\Delta t| < 3$ ns between the MBTS trigger counters on either side of the interaction point to suppress non-collision backgrounds. A coincidence between the ZDCs at forward and backward pseudorapidity is required to reject a variety of background processes, while maintaining high efficiency for non-Coulomb processes. Events satisfying these conditions are required to have a reconstructed primary vertex with $z_{\mathrm{vtx}}$ within 150~mm of the nominal centre of the ATLAS detector. About 48 million events pass the requirements for the analysis.

The Pb+Pb event centrality is characterized using the total transverse energy ($\Sigma \eT$) deposited in the FCal over the pseudorapidity range $3.2 < |\eta| < 4.9$ measured at the electromagnetic energy scale~\cite{Aad:2011he}. A larger $\Sigma \eT$ value corresponds to a more central collision. From an analysis of the $\Sigma \eT$ distribution after all trigger and event selections, the sampled fraction of the total inelastic cross section is estimated to be (98$\pm$2)\%~\cite{Aad:2011yr}.  The uncertainty associated with the centrality definition is evaluated by varying the effect of the trigger, event selection and background rejection requirements in the most peripheral FCal $\Sigma \eT$ interval~\cite{Aad:2011yr}. The FCal $\Sigma \eT$ distribution is divided into a set of 5\%-wide percentile bins, together with five 1\%-wide bins for the most central 5\% of the events.  A centrality interval refers to a percentile range, starting at 0\% for the most central collisions. Thus the 0-1\% centrality interval corresponds to the most central 1\% of the events; the 95-100\% centrality interval corresponds to the least central (i.e. most peripheral) 5\% of the events. A standard Glauber model Monte Carlo analysis is used to estimate the average number of participating nucleons, $\langle N_{\mathrm{part}}\rangle$, for each centrality interval~\cite{Miller:2007ri,Aad:2011yr}. These numbers are summarized in table~\ref{tab:cent}.

\begin{table}[h!]
\centering
\begin{tabular}{|l|c|c|c|c|c|}\hline
Centrality               & 0--1\%        & 1--2\%        & 2--3\%       & 3--4\%      &4--5\%       \tabularnewline\hline
$\langle N_{\mathrm{part}}\rangle$ & $400.6\pm1.3$& $392.6\pm1.8$&$383.2\pm2.1$ &$372.6\pm2.3$ &$361.8\pm2.5$\tabularnewline\hline
Centrality               & 0--5\%        & 5--10\%       & 10--15\%      &15--20\%   & 20--25\%       \tabularnewline\hline
$\langle N_{\mathrm{part}}\rangle$ & $382.2\pm2.0$&$330.3\pm3.0$ &$281.9\pm3.5$ &$239.5\pm3.8$& $202.6\pm3.9$\tabularnewline\hline
Centrality               & 25--30\%      & 30--35\%      & 35--40\%      &40--45\% & 45--50\%           \tabularnewline\hline
$\langle N_{\mathrm{part}}\rangle$ & $170.2\pm4.0$&$141.7\pm3.9$ &$116.8\pm3.8$ &$95.0\pm3.7$& $76.1\pm3.5$\tabularnewline\hline
Centrality               & 50--55\%      & 55--60\%      & 60--65\%      &65--70\%     &  \tabularnewline\hline
$\langle N_{\mathrm{part}}\rangle$ & $59.9\pm3.3$  &$46.1\pm3.0$  &$34.7\pm2.7$  &$25.4\pm2.3$&\tabularnewline\hline\hline
\end{tabular}
\caption{\label{tab:cent} The relationship between centrality intervals used in this paper and $\langle N_{\mathrm{part}}\rangle$ estimated from the Glauber model~\cite{Aad:2011yr}.}
\end{table}

The $v_n$ coefficients are measured using tracks reconstructed in the ID that have $\pT>0.5$~GeV and $|\eta|<2.5$. To improve the robustness of track reconstruction in the high-multiplicity environment of heavy ion collisions, more stringent requirements on track quality, compared to those defined for proton--proton collisions~\cite{Aad:2010ac}, are used. At least 9 hits in the silicon detectors (compared to a typical value of 11) are required for each track, with no missing Pixel hits and not more than 1 missing SCT hit, after taking into account the known non-operational modules. In addition, at its point of closest approach the track is required to be within 1~mm of the primary vertex in both the transverse and longitudinal directions~\cite{ATLAS:2011ah}. 

The efficiency, $\epsilon(\pT,\eta)$, of the track reconstruction and track selection cuts is evaluated using Pb+Pb Monte Carlo events produced with the HIJING event generator~\cite{Gyulassy:1994ew}. The generated particles in each event are rotated in azimuthal angle according to the procedure described in Ref.~\cite{Masera:2009zz} to give harmonic flow consistent with previous ATLAS measurements~\cite{ATLAS:2011ah,Aad:2012bu}. The response of the detector is simulated using GEANT4~\cite{Agostinelli:2002hh} and the resulting events are reconstructed with the same algorithms as applied to the data. The absolute efficiency increases with $\pT$ by 7\% between 0.5 GeV and 0.8 GeV, and varies only weakly for $\pT > 0.8$~GeV. However, the efficiency varies more strongly with $\eta$ and event multiplicity~\cite{trk}. For $\pT > 0.8$~GeV, it ranges from 72\% at $\eta = 0$ to 51\% for $|\eta| > 2$ in peripheral collisions, while it ranges from 72\% at $\eta = 0$ to about 42\% for $|\eta| > 2$ in central collisions. The fractional change of efficiency from most central to most peripheral collisions, when integrated over $\eta$ and $\pT$, is about 13\%. Contributions of fake tracks from random combinations of hits are generally negligible, reaching only 0.1\% for $|\eta|<1$ for the highest multiplicity events. This rate increases slightly at large $|\eta|$.
\section{Method and data analysis}
\label{sec:m1}
\begin{figure}[!b]
\centering
\includegraphics[width=1\columnwidth]{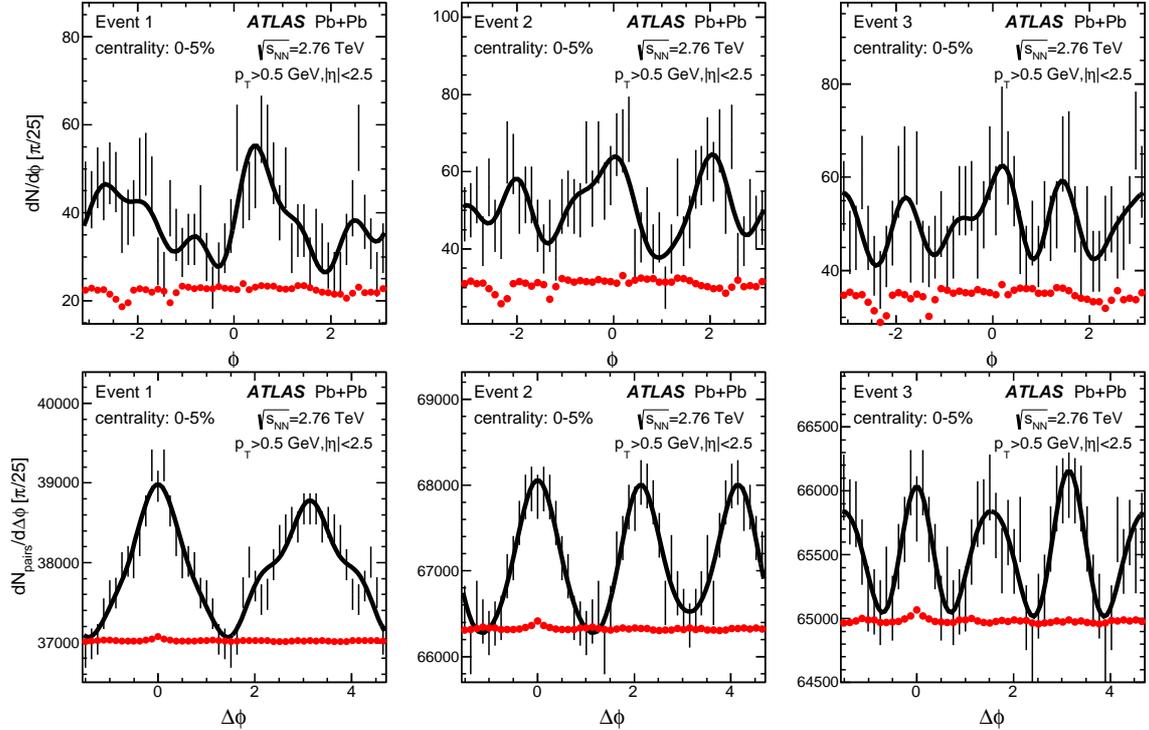}
\caption{\label{fig:evts} Single-track $\phi$ (top) and track-pair $\Delta\phi$ (bottom) distributions for three typical events (from left to right) in the 0--5\% centrality interval. The pair distributions are folded into the $[-0.5\pi,1.5\pi]$ interval. The bars indicate the statistical uncertainties of the foreground distributions, the solid curves indicate a Fourier parameterization including the first six harmonics: ${\rm d}N/{\rm d}\phi=A(1+2\sum_{i=1}^{6}c_n\cos n(\phi-\Psi_n))$ for single-track distributions and ${\rm d}N/{\rm d}\Delta\phi=A(1+2\sum_{i=1}^{6}c_n\cos n(\Delta\phi))$ for track-pair distributions, and the solid points indicate the event-averaged distributions (arbitrary normalization).}
\end{figure}

To illustrate the level of EbyE fluctuations in the data, the top panels of figure~\ref{fig:evts} show the azimuthal distribution of charged particle tracks with $\pT>0.5$~GeV for three typical events in the 0-5\% centrality interval. The corresponding track-pair $\Delta\phi$ distributions from the same events are shown in the bottom panels. For each pair of particles two $\Delta\phi$ entries, $|\phi_1-\phi_2|$ and $-|\phi_1-\phi_2|$, are made each with a weight of 1/2, and then folded into the $[-0.5\pi,1.5\pi]$ interval. Rich EbyE patterns, beyond the structures in the event-averaged distributions shown by the solid points (arbitrary normalization), are observed. These EbyE distributions are the inputs to the EbyE $v_n$ analyses.

The azimuthal distribution of charged particles in an event is written as a Fourier series, as in eq.~(\ref{eq:flow}):
\begin{eqnarray}
\label{eq:1}
&&\hspace*{-0.2cm}\frac{{\rm d}N}{{\rm d}\phi}\propto 1+2\sum_{n=1}^{\infty}v_n^{\mathrm{obs}}\cos n(\phi-\Psi_{n}^{\mathrm{obs}})=  1+2\sum_{n=1}^{\infty}\left(v_{n,{x}}^{\mathrm{obs}}\cos n\phi+v_{n,{y}}^{\mathrm{obs}}\sin n\phi\right)\;,\\\nonumber
&&\hspace*{-0.2cm}v_n^{\mathrm{obs}} = \sqrt{\left(v_{n,{x}}^{\mathrm{obs}}\right)^2+\left(v_{n,{y}}^{\mathrm{obs}}\right)^2},v_{n,{x}}^{\mathrm{obs}} = v_n^{\mathrm{obs}}\cos n\Psi_{n}^{\mathrm{obs}}=\langle\cos n\phi\rangle,v_{n,{y}}^{\mathrm{obs}} =  v_n^{\mathrm{obs}}\sin n\Psi_{n}^{\mathrm{obs}}= \langle\sin n\phi\rangle,\\
\end{eqnarray}
where the averages are over all particles in the event for the required $\eta$ range. The $v_n^{\mathrm{obs}}$ is the magnitude of the observed EbyE per-particle flow vector: $\overrightharp{v}_n^{\;\mathrm{obs}}=(v_{n,{x}}^{\mathrm{obs}},v_{n,{y}}^{\mathrm{obs}})$. In the limit of very large multiplicity and in the absence of non-flow effects, it approaches the true flow signal: $v_n^{\mathrm{obs}}\rightarrow v_n$. The key issue in measuring the EbyE $v_n$ is to determine the response function $p(v_n^{\mathrm{obs}}|v_n)$, which can be used to unfold the smearing effect due to the finite number of detected particles. Possible non-flow effects from short-range correlations, such as resonance decays, Bose--Einstein correlations and jets, also need to be suppressed.

The rest of this section sets out the steps to obtain the unfolded $v_n$ distribution. Since the data-driven unfolding technique has rarely been used in the study of flow phenomena, details are provided to facilitate the understanding of the methods and systematic uncertainties. Section~\ref{sec:m1.1} explains how $v_n^{\mathrm{obs}}$ and the associated response function can be obtained from the EbyE single-particle distributions, such as those shown in the top panels of figure~\ref{fig:evts}. Section~\ref{sec:m1.2} describes how $v_n^{\mathrm{obs}}$ and the response function can be obtained from EbyE two-particle correlations (2PC), similar to those shown in the lower panels of figure~\ref{fig:evts}. In this paper the 2PC approach is used primarily as a consistency check. The Bayesian unfolding procedure, applicable to either the single-particle or 2PC data, is described in section~\ref{sec:m1.3}. The performance of the unfolding is described in section~\ref{sec:m2.1}, while the systematic uncertainties are discussed in section~\ref{sec:m2.2}. 

\subsection{Single-particle method}
\label{sec:m1.1}
The azimuthal distribution of particles in figure~\ref{fig:evts} needs to be corrected for non-uniform detector acceptance. This is achieved by dividing the foreground distribution $(S)$ of a given event by the acceptance function $(B)$ obtained as the $\phi$ distribution of all tracks in all events (see top panels of figure~\ref{fig:evts}):
\begin{eqnarray}
\label{eq:2}
&& \frac{{\rm d}N}{{\rm d}\phi} \propto \frac{S(\phi)}{B(\phi)}=\frac{1+2\sum_{n=1}^{\infty}\left(v_{n,{x}}^{\mathrm{raw}}\cos n\phi+v_{n,{y}}^{\mathrm{raw}}\sin n\phi)\right)}{1+2\sum_{n=1}^{\infty}\left(v_{n,{x}}^{\mathrm{det}}\cos n\phi+v_{n,{y}}^{\mathrm{det}}\sin n\phi)\right)}\;,\\
&&v_{n,{x}}^{\mathrm{raw}}=\frac{\sum_{i} \left(\cos n\phi_i\right)/\epsilon(\eta_i,p_{\mathrm{T},i})}{\sum_{i} 1/\epsilon(\eta_i,p_{\mathrm{T},i})},\;\;\; v_{n,{y}}^{\mathrm{raw}}=\frac{\sum_{i} \left(\sin n\phi_i\right)/\epsilon(\eta_i,p_{\mathrm{T},i})}{\sum_{i} 1/\epsilon(\eta_i,p_{\mathrm{T},i})}\;,
\end{eqnarray}
where the index $i$ runs over all tracks in an event, $\epsilon(\eta,p_{\mathrm{T}})$ is the tracking efficiency for a given centrality interval, and $v_{n,{x}}^{\mathrm{det}}$ and $v_{n,{y}}^{\mathrm{det}}$ are Fourier coefficients of the acceptance function in azimuth, also weighted by the inverse of the tracking efficiency. The influence of the structures in the acceptance function can be accounted for by taking the leading-order term of the Taylor expansion of $1/B(\phi)$ in terms of $\cos n\phi$ and $\sin n\phi$:
\begin{eqnarray}
\label{eq:3}
v_{n,{x}}^{\mathrm{obs}}\approx v_{n,{x}}^{\mathrm{raw}}-v_{n,{x}}^{\mathrm{det}},\;\;\; v_{n,{y}}^{\mathrm{obs}}\approx v_{n,{y}}^{\mathrm{raw}}-v_{n,{y}}^{\mathrm{det}}\;,
\end{eqnarray}
where the values of $v_{n,{x} \mathrm{\;or\;} {y}}^{\mathrm{det}}$ are less than 0.007 for $n=2$--4. The higher-order corrections to eq.~(\ref{eq:3}) involve products of $v_{n,{x} \mathrm{\;or\;} {y}}^{\mathrm{raw}}$ and $v_{n,{x} \mathrm{\;or\;} {y}}^{\mathrm{det}}$. They have been estimated and found to have negligible impact on the final $v_n$ distributions for $n=2$--4.

\begin{figure}[!t]
\centering
\includegraphics[width=1\columnwidth]{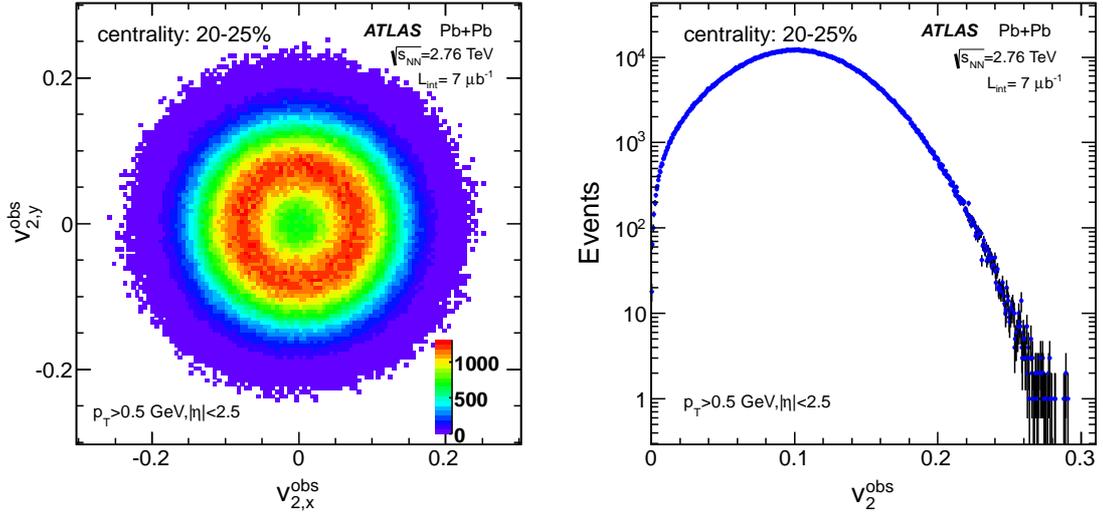}
\caption{\label{fig:idea1} The distribution of EbyE per-particle flow vector ${\protect\overrightharp v}_2^{\;\mathrm{obs}}$ (left panel) and its magnitude $v_2^{\mathrm{obs}}$ (right panel) for events in the 20--25\% centrality interval.}
\end{figure}

Figure~\ref{fig:idea1} shows the distribution of the EbyE per-particle flow vector $\overrightharp{v}_2^{\;\mathrm{obs}}=(v_{2,{x}}^{\mathrm{obs}},v_{2,{y}}^{\mathrm{obs}})$ and $v_2^{\mathrm{obs}}$ obtained for charged particles with $\pT>0.5$~GeV in the 20--25\% centrality interval. The azimuthal symmetry in the left panel reflects the random orientation of $\overrightharp{v}_2^{\;\mathrm{obs}}$ because of the random orientation of the impact parameter. Due to the finite track multiplicity, the measured flow vector is expected to be smeared randomly around the true flow vector by a 2D response function $p(\overrightharp{v}_n^{\;\mathrm{obs}}|\overrightharp{v}_n)$.
\begin{figure}[!h]
\centering
\includegraphics[width=1\columnwidth]{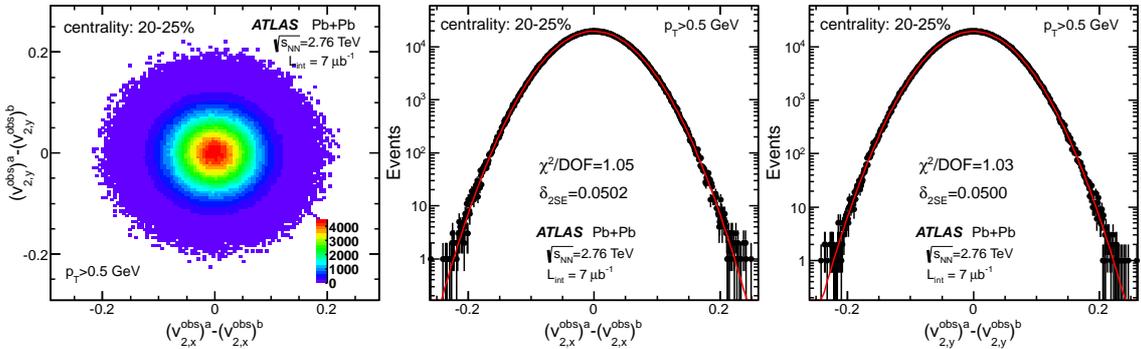}
\caption{\label{fig:idea2} Left: The distribution of the difference between the EbyE per-particle flow vectors of the two half-IDs for events in the 20--25\% centrality interval for $n=2$. Middle: The $x$-projection overlaid with a fit to a Gaussian. Right: The $y$-projection overlaid with a fit to a Gaussian. The width from the fit, $\delta_{_{\mathrm{2SE}}}$, and the quality of the fit, $\chi^2/$DOF, are also shown.}
\end{figure}

In order to determine $p(\overrightharp{v}_n^{\;\mathrm{obs}}|\overrightharp{v}_n)$, the tracks in the entire inner detector (referred to as full-ID) for a given event are divided into two subevents with symmetric $\eta$ range, $\eta>0$ and $\eta<0$ (labelled by $a$ and $b$ and referred to as half-ID). The two half-IDs have the same average track multiplicity to within 1\%. The distribution of flow vector differences between the two subevents, $p_{\mathrm{sub}}\left((\overrightharp{v}_n^{\;\mathrm{obs}})^{\rm a}-(\overrightharp{v}_n^{\;\mathrm{obs}})^{\rm b}\right)$, is then obtained and is shown in the left panel of figure~\ref{fig:idea2}. The physical flow signal cancels in this distribution such that it contains mainly the effects of statistical smearing and non-flow. The middle and right panels of figure~\ref{fig:idea2} show the $x$- and $y$- projections of the distribution, together with fits to a Gaussian function. The fits describe the data very well ($\chi^2/$DOF$\approx1$) across five orders of magnitude with the same widths in both directions, implying that the smearing effects and any effects due to non-flow short-range correlations are purely statistical. This would be the case if either the non-flow effects are small and the smearing is mostly driven by finite particle multiplicity, or the number of sources responsible for non-flow is proportional to the multiplicity and they are not correlated between the subevents~\cite{Jia:2013tja}. The latter could be true for resonance decays, Bose--Einstein correlations, and jets. Similar behaviour is observed for all harmonics up to centrality interval 50--55\%. Beyond that the distributions are found to be described better by the Student's t-distribution, which is a general probability density function for the difference between two estimates of the mean from independent samples. The t-distribution approaches a Gaussian distribution when the number of tracks is large.

Denoting the width of these 1D distributions by $\delta_{_{\mathrm{2SE}}}$, the widths of the response functions for the half-ID and the full-ID are $\delta_{_{\mathrm{2SE}}}/\sqrt{2}$ and $\delta_{_{\mathrm{2SE}}}/2$, respectively. The response functions themselves can be obtained by rescaling the left panel of figure~\ref{fig:idea2} in both dimensions by constant factors of 2 and $\sqrt{2}$ for the full-ID and half-ID, respectively~\cite{Jia:2013tja}: 
\begin{eqnarray}
p(\overrightharp{v}_n^{\;\mathrm{obs}}|\overrightharp{v}_n)\propto e^{-\frac{\left(\overrightharp{v}^{\mathrm{\;obs}}_n-\overrightharp{v}_n\right)^2}{2\delta^2}},\;\;\;\delta= &\left\{\begin{array}{ll}
  \delta_{_{\mathrm{2SE}}}/\sqrt{2}  &\textrm{ for half-ID } 
    \\
   \delta_{_{\mathrm{2SE}}}/2  &\textrm{ for full-ID } 
    \end{array}\right.,\nonumber\,\\\label{eq:5bb}
\end{eqnarray}
This scaling behaviour was found to be valid in a Monte-Carlo study based on the HIJING event generator~\cite{Jia:2013tja}.
Integrated over azimuth, Eq.~\ref{eq:5bb} reduces to a Bessel--Gaussian function in 1D:
\begin{eqnarray}
\label{eq:5b}
p(v_n^{\mathrm{obs}}|v_n)\propto v_n^{\mathrm{obs}}e^{-\frac{(v_n^{\mathrm{obs}})^2+v_n^2}{2\delta^2}} I_0\left(\frac{v_n^{\mathrm{obs}}v_n}{\delta^2}\right)\;.
\end{eqnarray}
The difference between the observed and the true signal, denoted by $s=v_n^{\mathrm{obs}}-v_n$, accounts for the statistical smearing. The similarity between eq.~(\ref{eq:5b}) and eq.~(\ref{eq:fluc2}) is a direct consequence of the 2D Gaussian smearing. However, the smearing leading to eq.~(\ref{eq:5b}) is due to the finite charge-particle multiplicity, while the smearing leading to eq.~(\ref{eq:fluc2}) is due to the intrinsic flow fluctuations associated with the initial geometry. Hence the smearing in eq.~(\ref{eq:5b}) is expected to increase the observed $\delta_{_{v_n}}$ value but the value of $v_n^{\mathrm{RP}}$ should be relatively stable.

The analytical expression eq.~(\ref{eq:5b}) can be used to unfold the $v_n^{\mathrm{obs}}$ distribution, such as that shown in the right panel of figure~\ref{fig:idea1}. Alternatively, the measured distribution $p(v_n^{\mathrm{obs}}|v_n)$, taking into account sample statistics, can be used directly in the unfolding. This measured distribution is obtained by integrating out the azimuthal angle in the 2D response function, and the response function is obtained by rescaling the left panel of figure~\ref{fig:idea2} as described earlier. This approach is more suitable for peripheral collisions where the analytical description using eq.~(\ref{eq:5b}) is not precise enough.

\subsection{Two-particle correlation method}
\label{sec:m1.2}
The EbyE two-particle correlation (2PC) method starts from the $\Delta\phi$ information in each event, where $\Delta\phi$ is calculated for each pair of charged tracks as described at the start of section~\ref{sec:m1}. In order to reduce the effect of short-range correlations in $\eta$, the tracks in each pair are taken from different half-IDs. This procedure corresponds to convolving the azimuthal distributions of single particles in the two half-IDs:
\begin{eqnarray}
\nonumber
\frac{{\rm d}N}{{\rm d}\Delta\phi} &\propto& \left[1+2\sum_n\left(v_{n,{x}}^{\mathrm{obs_{\rm a}}}\cos n\phi_{\rm a}+ v_{n,{y}}^{\mathrm{obs_{\rm a}}}\sin n\phi_{\rm a}\right)\right]\otimes\left[1+2\sum_n\left(v_{n,{x}}^{\mathrm{obs_{\rm b}}}\cos n\phi_{\rm b}+ v_{n,{y}}^{\mathrm{obs_{\rm b}}}\sin n\phi_{\rm b}\right)\right]\\\nonumber
&=&1+2\sum_n\left[\left(v_{n,{x}}^{\mathrm{obs_{\rm a}}}v_{n,{x}}^{\mathrm{obs_{\rm b}}}+v_{n,{y}}^{\mathrm{obs_{\rm a}}}v_{n,{y}}^{\mathrm{obs_{\rm b}}}\right)\cos n\Delta\phi+ \left(v_{n,{x}}^{\mathrm{obs_{\rm a}}}v_{n,{y}}^{\mathrm{obs_{\rm b}}}-v_{n,{y}}^{\mathrm{obs_{\rm a}}}v_{n,{x}}^{\mathrm{obs_{\rm b}}}\right)\sin n\Delta\phi\right]\\\label{eq:p1}
&\equiv& 1+2\sum_n\left(A_n\cos n\Delta\phi+ B_n\sin n\Delta\phi\right)\;,
\end{eqnarray}
where $A_n=\langle\cos n\Delta\phi\rangle$ and $B_n=\langle\sin n\Delta\phi\rangle$. The parameters $A_n$ and $B_n$ are calculated by averaging over the pairs in each event, with each track weighted by the tracking efficiency, as in eq.~(\ref{eq:2}). Due to a large rapidity gap on average between the two particles in each pair, the non-flow effects in eq.~\ref{eq:p1} are naturally suppressed compared with the single particle distribution of eq.~\ref{eq:1}.

An EbyE track-pair variable $v_{n,n}^{\mathrm{obs}}$ is subsequently calculated for each event:
\begin{align}
\label{eq:p2}
v_{n,n}^{\mathrm{obs}} \equiv \sqrt{A_n^2+B_n^2} = \sqrt{\left[\left(v_{n,{x}}^{\mathrm{obs_{\rm a}}}\right)^2+\left(v_{n,{y}}^{\mathrm{obs_{\rm a}}}\right)^2\right]\left[\left(v_{n,{x}}^{\mathrm{obs_{\rm b}}}\right)^2+\left(v_{n,{y}}^{\mathrm{obs_{\rm b}}}\right)^2\right]}=v_n^{\mathrm{obs_{\rm a}}}v_n^{\mathrm{obs_{\rm b}}}\;.
\end{align}
The observed flow signal from the two-particle correlation analysis is then calculated as:
\begin{align}
\label{eq:p3}
v_n^{\mathrm{obs,2PC}} \equiv \sqrt{v_{n,n}^{\mathrm{obs}}} = \sqrt{v_n^{\mathrm{obs_{\rm a}}}v_n^{\mathrm{obs_{\rm b}}}}= \sqrt{(v_n+s_{\rm a})(v_n+s_{\rm b})}\;,
\end{align}
where $s_{\rm a}=v_n^{\mathrm{obs_{\rm a}}}-v_n$ and $s_{\rm b}=v_n^{\mathrm{obs_{\rm b}}}-v_n$ are independent variables described by the probability distribution in eq.~(\ref{eq:5b}) with $\delta=\delta_{_{\mathrm{2SE}}}/\sqrt{2}$. The response function for $v_n^{\mathrm{obs,2PC}}$ is very different from that for the single-particle method, but nevertheless can be either calculated analytically via eq.~(\ref{eq:5b}) or generated from the measured distribution such as that shown in figure~\ref{fig:idea2}. For small $v_n$ values, the $s_{\rm a}s_{\rm b}$ term dominates eq.~(\ref{eq:p3}) and the distribution of $v_n^{\mathrm{obs,2PC}}$ is broader than $v_n^{\mathrm{obs}}$. For large $v_n$ values, the distributions of $s_{\rm a}$ and $s_{\rm b}$ are approximately described by Gaussian functions and hence:
\begin{align}
\label{eq:p4}
v_n^{\mathrm{obs,2PC}} \approx \sqrt{v_n^2+v_n(s_{\rm a}+s_{\rm b})}\approx v_n+\frac{s_{\rm a}+s_{\rm b}}{2}\equiv v_n+s\;,
\end{align}
where the fact that the average of two Gaussian random variables is equivalent to a Gaussian with a narrower width has been used, and the smearing of the flow vector for the half-IDs ($s_{\rm a}$ and $s_{\rm b}$) is taken to be a factor of $\sqrt{2}$ broader than that for the full-ID ($s$). Hence the distribution of $v_n^{\mathrm{obs,2PC}}$ is expected to approach the $v_n^{\mathrm{obs}}$ distribution of the full-ID when $v_n\gg\delta_{_{\mathrm{2SE}}}/\sqrt{2}$.

\subsection{Unfolding procedure}
\label{sec:m1.3}
In this analysis, the standard Bayesian unfolding procedure~\cite{Agostini}, as implemented in the RooUnfold framework~\cite{unfold}, is used to obtain the $v_n$ distribution. In this procedure, the true $v_n$ distribution (``cause'' $\hat{c}$) is obtained from the measured $v_n^{\mathrm{obs}}$ or $v_n^{\mathrm{obs,2PC}}$ distribution (``effect'' $\hat{e}$) and the response function $A_{ji}\equiv p(e_j|c_i)$ ($i$,$j$ are bins) as:
\begin{align}
\label{eq:bay1}
\hat{c}^{\mathrm{iter}+1} = \hat{M}^{\mathrm{iter}}\hat{e}, \;\;\; M_{ij}^{\mathrm{iter}} = \frac{A_{ji}c_i^{\mathrm{iter}}}{\sum_{m,k}A_{mi}A_{jk}c_k^{\mathrm{iter}}}\;,
\end{align}
where the unfolding matrix $\hat{M}^{0}$ is determined from the response function and some initial estimate of the true distribution $\hat{c}^0$ (referred to as the prior). The matrix $\hat{M}^{0}$ is used to obtain the unfolded distribution $\hat{c}^{1}$ and $\hat{M}^{1}$, and the process is then iterated. More iterations reduce the dependence on the prior and give results closer to the true distributions but with increased statistical fluctuations. Therefore the number of iterations $N_{\mathrm{iter}}$ is adjusted according to the sample size and binning. The prior can be chosen to be the $v_{n}^{\mathrm{obs}}$ distribution from the full-ID for the single-particle unfolding, or the $v_{n}^{\mathrm{obs,2PC}}$ distribution obtained by convolving the two half-IDs (eq.~(\ref{eq:p1})) for the 2PC unfolding. However, a more realistic prior can be obtained by rescaling the $v_n^{\mathrm{obs}}$ in each event by a constant factor $R$ to obtain a distribution with a mean that is closer to that of the true distribution:
\begin{eqnarray}
\nonumber
v_n^{\mathrm{obs}} \rightarrow Rv_n^{\mathrm{obs}}, \;\;R= \frac{v_n^{\mathrm{EP}}}{\langle v_n^{\mathrm{obs}}\rangle}\frac{1}{1+\left(\sqrt{1+\left(\sigma_{v_n}^{\mathrm{obs}}/\langle v_n^{\mathrm{obs}}\rangle\right)^2}-1\right)f}, \;\; f = 0, 0.5, 1, 1.5, 2, 2.5\;,\\\label{eq:pri0}
\end{eqnarray}
where $\langle v_n^{\mathrm{obs}}\rangle$ and $\sigma_{v_n}^{\mathrm{obs}}$ are the mean and the standard deviation of the $v_n^{\mathrm{obs}}$ distribution, respectively, and $v_n^{\mathrm{EP}}$ is measured using the FCal event plane method from Ref.~\cite{Aad:2012bu} with the same dataset and the same track selection criteria. The EP method is known to measure a value between the mean and the root-mean-square of the true $v_n$~\cite{Alver:2008zza,Ollitrault:2009ie} (see figure~\ref{fig:result3a}):
\begin{eqnarray}
\label{eq:pri1}
\langle v_n\rangle\leq v_n^{\mathrm{EP}}\leq\sqrt{\langle v_n^2\rangle}=\sqrt{\langle v_n\rangle^2+\sigma_{v_n}^2}\;. 
\end{eqnarray}
The lower limit is reached when the resolution factor~\cite{Aad:2012bu} used in the EP method approaches one, and the upper limit is reached when the resolution factor is close to zero. Thus $f=0$ (default choice) gives a prior that is slightly broader than the true distribution, $f=1$ gives a distribution that has a mean close to the true distribution, and $f>1$ typically gives a distribution that is narrower than the true distribution.

The unfolding procedure in this analysis has several beneficial features:
\begin{enumerate}
\item
 The response function is obtained entirely from the data using the subevent method described above (eq.~(\ref{eq:5bb})).
\item Each event provides one entry for the $v_n^{\mathrm{obs}}$ distribution and the response function (no efficiency loss), and these distributions can be determined with high precision (from about 2.4 million events for each 5\% centrality interval).
\end{enumerate}

\subsection{Unfolding performance}
\label{sec:m2.1}
This section describes the unfolding based on the single-particle method and summarizes a series of checks used to verify the robustness of the results: a) the number of iterations used, b) comparison with results obtained from a smaller $\eta$ range, c) variation of the priors, d) comparison with the results obtained using the 2PC method, and e) estimation of possible biases due to short-range correlations by varying the $\eta$ gap between the two half-IDs. Only a small subset of the available figures is presented here; a complementary selection can be found in appendix A.

\begin{figure}[!h]
\centering
\includegraphics[width=1\columnwidth]{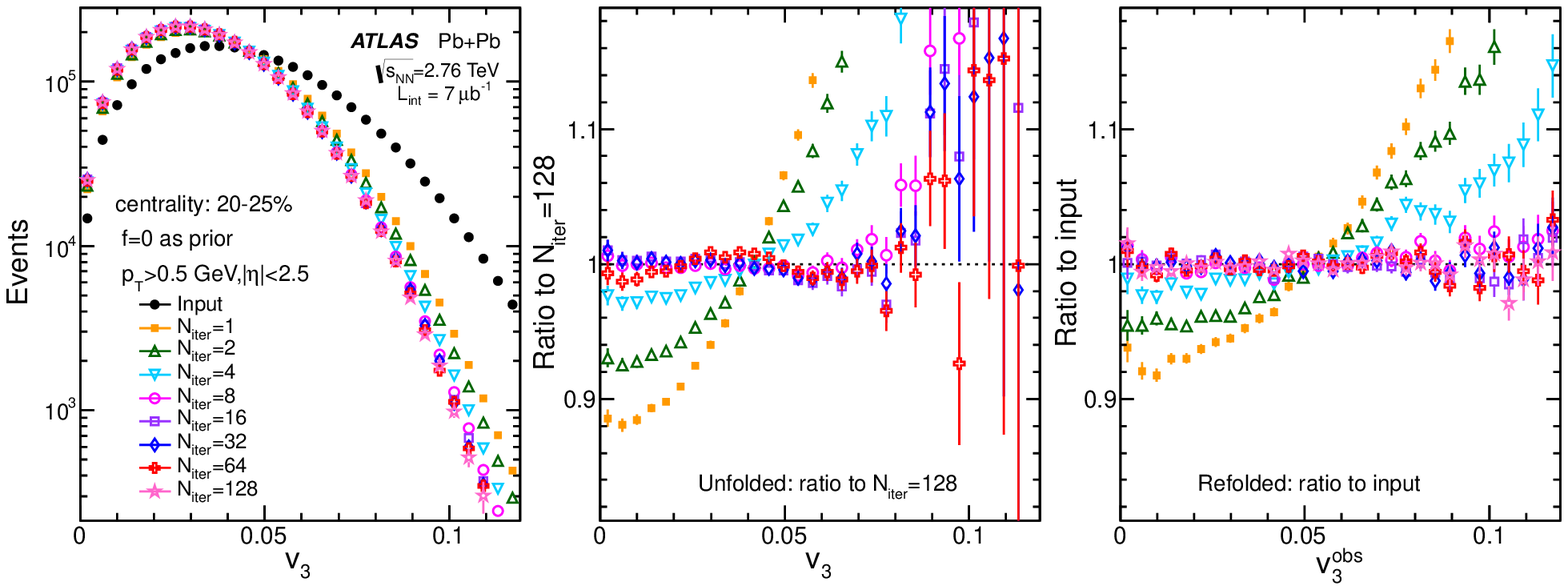}\vspace*{-0.3cm}
\caption{\label{fig:perf2} The performance of the unfolding of $v_3$ for the 20--25\% centrality interval (left panel) for various $N_{\mathrm{iter}}$, the ratios of the unfolded distributions to the results after 128 iterations (middle panel), and the ratios of the refolded distributions to the input $v_3^{\mathrm{obs}}$ (right panel).}
\end{figure}

\begin{figure}[!h]
\centering
\includegraphics[width=1\columnwidth]{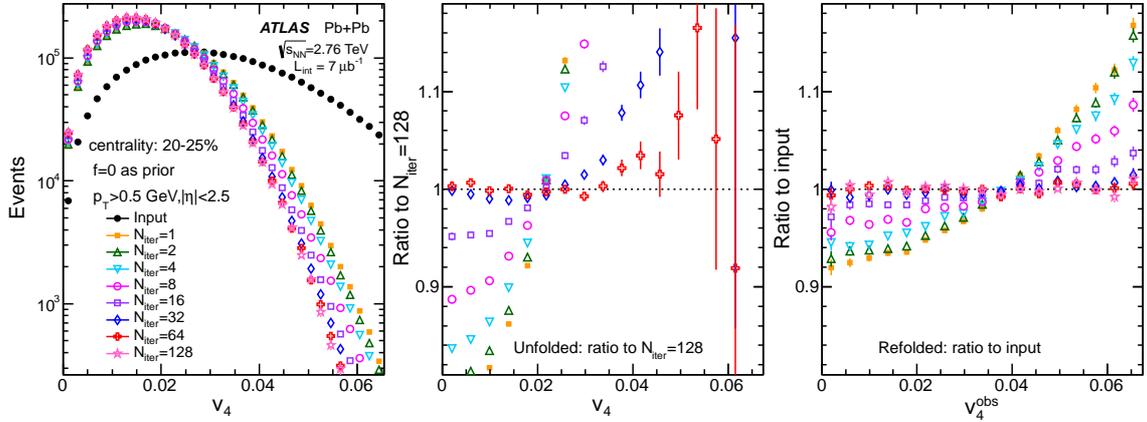}\vspace*{-0.3cm}
\caption{\label{fig:perf3} The performance of the unfolding of $v_4$ for the 20--25\% centrality interval (left panel) for various $N_{\mathrm{iter}}$, the ratios of the unfolded distributions to the results after 128 iterations (middle panel), and the ratios of the refolded distributions to the input $v_4^{\mathrm{obs}}$ (right panel).}
\end{figure}

The left and middle panels of figure~\ref{fig:perf1} show the convergence behaviour of the unfolding based on the single-particle method for $v_2$ in the 20--25\% centrality interval measured with the full-ID. Around the peak of the distribution, the results converge to within a few percent of the final answer by $N_{\mathrm{iter}}=4$, but the convergence is slower in the tails and there are small, systematic improvements at the level of a few percent for $N_{\mathrm{iter}}\ge8$. The refolded distributions (right panel), obtained by convolving the unfolded distributions with the response function, indicate that convergence is reached for $N_{\mathrm{iter}}\ge8$. Figures~\ref{fig:perf2} and \ref{fig:perf3} show similar distributions for $v_3$ and $v_4$. The performance of the unfolding is similar to that shown in figure~\ref{fig:perf1}, except that the tails of the unfolded distributions require more iterations to converge. For example, figure~\ref{fig:perf3} suggests that the bulk region of the $v_4$ distributions has converged by $N_{\mathrm{iter}}=32$, but the tails still exhibit some small changes up to $N_{\mathrm{iter}}=64$. The slower convergence for higher-order harmonics reflects the fact that the physical $v_n$ signal is smaller for larger $n$, while the values of $\delta_{_{\mathrm{2SE}}}$ change only weakly with $n$. These studies are repeated for all centrality intervals. In general, more iterations are needed for peripheral collisions due to the increase in $\delta_{_{\mathrm{2SE}}}$.
\begin{figure}[!b]
\centering
\includegraphics[width=1\columnwidth]{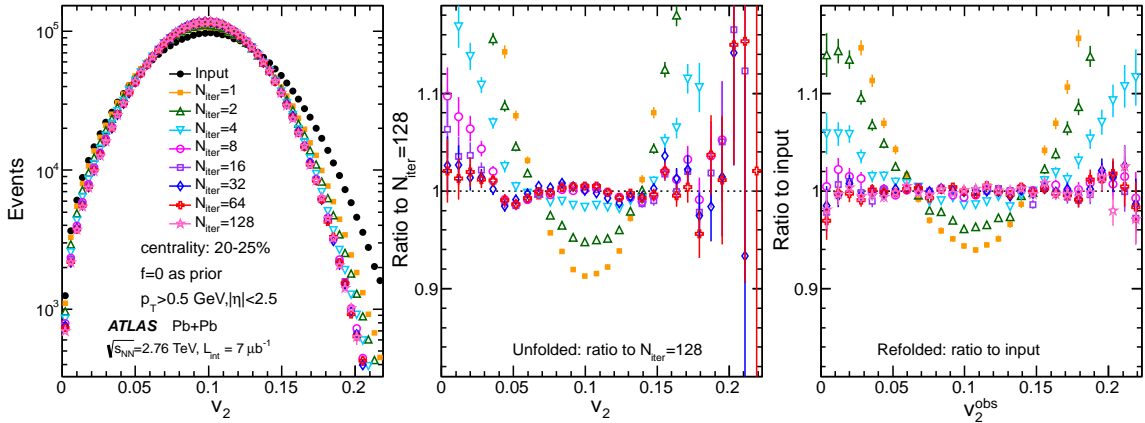}\vspace*{-0.3cm}
\caption{\label{fig:perf1} The performance of the unfolding of $v_2$ for the 20--25\% centrality interval (left panel) for various $N_{\mathrm{iter}}$, the ratios of the unfolded distributions to the results after 128 iterations (middle panel), and the ratios of the refolded distributions to the input $v_2^{\mathrm{obs}}$ (right panel).}
\end{figure}

The statistical uncertainties in the unfolding procedure are verified via a resampling technique~\cite{resample}. For small $N_{\mathrm{iter}}$, the statistical uncertainties as given by the diagonal entries of the covariance matrix are much smaller than $\sqrt{N}$, where $N$ is the number of entries in each bin, indicating the presence of statistical bias in the prior. However, these uncertainties increase with $N_{\mathrm{iter}}$, and generally approach $\sqrt{N}$ for $64\leq N_{\mathrm{iter}}\leq128$. In this analysis, the centrality range for each harmonic is chosen such that the difference between $N_{\mathrm{iter}}=32$ and $N_{\mathrm{iter}}=128$ is less than 10\%. The centrality ranges are 0--70\% for $v_2$, 0--60\% for $v_3$ and 0--45\% for $v_4$.

\begin{figure}[!t]
\centering
\includegraphics[width=1\columnwidth]{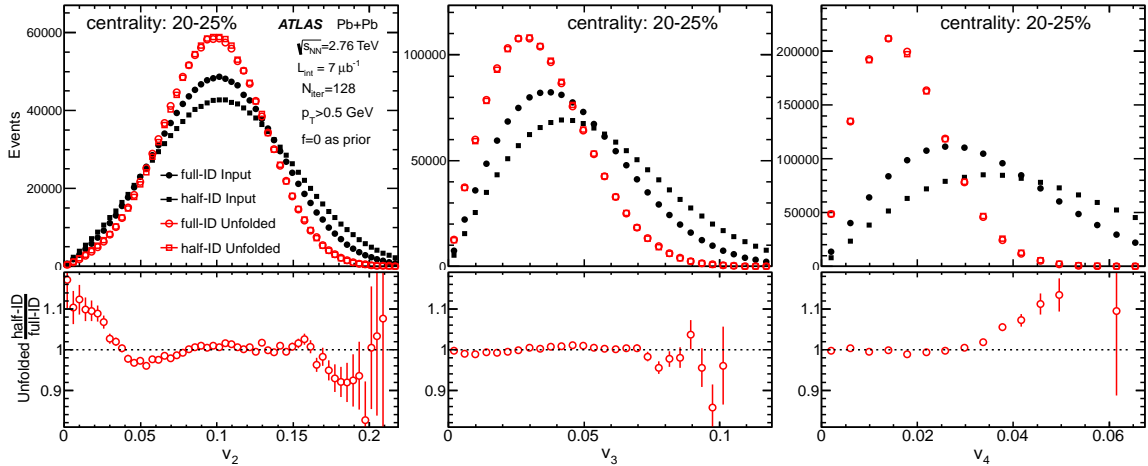}\vspace*{-0.3cm}
\caption{\label{fig:perf4} Comparison of the input distributions (solid symbols) and unfolded distributions for $N_{\mathrm{iter}}= 128$ (open symbols) between the half-ID and the full-ID in the 20--25\% centrality interval. The ratios of half-ID to full-ID unfolded results are shown in the bottom panels. The results are shown for $v_2$ (left panels), $v_3$ (middle panels) and $v_4$ (right panels).}
\end{figure}

The robustness of the unfolding procedure is checked by comparing the results measured independently for the half-ID and the full-ID. The results are shown in figure~\ref{fig:perf4}. Despite the large differences between their initial distributions, the final unfolded results agree to within a few percent in the bulk region of the unfolded distribution, and they are nearly indistinguishable on a linear scale. This agreement also implies that the influence due to the slight difference (up to 1\%) in the average track multiplicity between the two subevents is small. Systematic differences are observed in the tails of the distributions for $v_4$, especially in peripheral collisions, where the half-ID results are slightly broader. This behaviour reflects mainly the deviation from the expected truth (residual non-convergence) for the half-ID unfolding, since the response function is a factor of $\sqrt{2}$ broader than that for the full-ID.

\begin{figure}[!t]
\centering
\includegraphics[width=1\columnwidth]{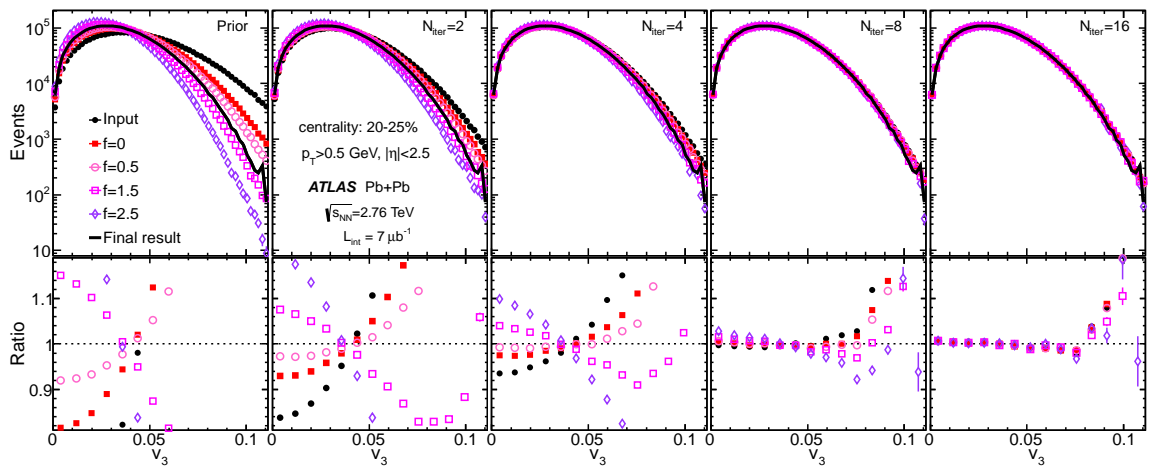}\vspace*{-0.2cm}
\caption{\label{fig:perf5} Convergence behaviour of $v_3$ in the 20--25\% centrality interval for five choices of priors for different $N_{\mathrm{iter}}$ from left to right. The top panels show the distributions after a certain number of iterations and bottom panels show the ratios to the result for $N_{\mathrm{iter}}=128$. A common reference, shown by the solid lines in the top panels, is calculated by averaging the results for $f=0$ and $f=0.5$ with $N_{\mathrm{iter}}=128$. }
\end{figure}

A wide range of priors has been tried in this analysis, consisting of the measured $v_n^{\mathrm{obs}}$ distribution and the six rescaled distributions defined by eq.~(\ref{eq:pri0}). Figure~\ref{fig:perf5} compares the convergence behaviour of these priors for $v_3$ in the 20--25\% centrality interval. Despite the significantly different initial distributions, the unfolded distributions converge to the same answer, to within a few percent, for $N_{\mathrm{iter}}\ge16$. A prior that is narrower than the unfolded distribution leads to convergence in one direction, and a broader prior leads to convergence from the other direction. Taken together, these checks allow a quantitative evaluation of the residual non-convergence in the unfolded distributions.

\begin{figure}[!t]
\centering
\includegraphics[width=1\columnwidth]{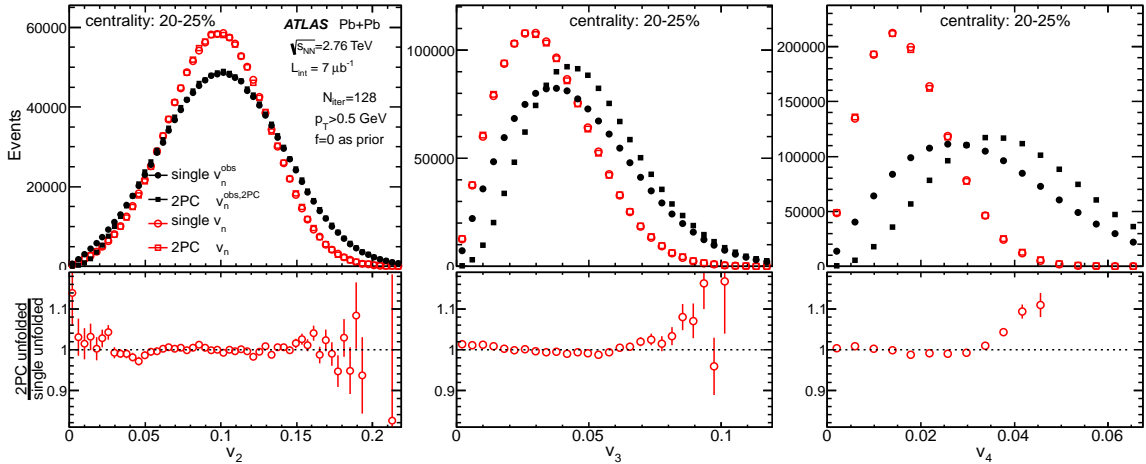}\vspace*{-0.3cm}
\caption{\label{fig:perf6}  Comparison of the input distributions (solid symbols) and unfolded distributions for $N_{\mathrm{iter}}= 128$ (open symbols) between the single-particle unfolding and 2PC unfolding in the 20--25\% centrality interval for $v_2$ (left panels), $v_3$ (middle panels) and $v_4$ (right panels). The ratios of 2PC to single-particle unfolded results are shown in the bottom panels.}
\end{figure}

Figure~\ref{fig:perf6} compares the convergence behaviour between unfolding of single-particle $v_n^{\mathrm{obs}}$ and unfolding of $v_n^{\mathrm{obs,2PC}}$ in the 20--25\% centrality interval. Despite very different response functions and initial distributions, the unfolded results agree with each other to within a few percent in the bulk region of the unfolded distribution. The systematic deviations in the tails of the $v_4$ distribution (bottom-right panel) are due mainly to the remaining non-convergence in the 2PC method, which has a broader response function than the single-particle method.

One important issue in the EbyE $v_n$ study is the extent to which the unfolded results are biased by non-flow short-range correlations, which may influence both the $v_n^{\mathrm{obs}}$ distributions and the response functions. This influence contributes to both the $v_n^{\mathrm{obs}}=|\protect\overrightharp{v}_n^{\;\mathrm{obs}}|$ distributions and response functions obtained from ${(\overrightharp{v}_n^{\;\mathrm{obs}})^{\rm a}-(\overrightharp{v}_n^{\;\mathrm{obs}})^{\rm b}}$ (figure~\ref{fig:idea2}), and hence are expected largely to cancel out in the unfolding procedure. This conclusion is supported by a detailed Monte-Carlo model study based on the HIJING event generator with a realistic flow afterburner~\cite{Jia:2013tja}, where the unfolding performance was evaluated. It is also supported by the consistency between the single-particle and 2PC methods (figure~\ref{fig:perf6}), which have different sensitivities to the non-flow effects. Furthermore, both unfolding methods have been repeated requiring a minimum $\eta$ gap between the two subevents used to obtain the input distributions and the response functions. Six additional cases, requiring $\eta_{\mathrm{gap}}=0.2,0.4,0.6,0.8,1.0,2.0$, have been studied and the results have been compared (see figure~\ref{fig:perf6a}). The unfolded $v_n$ distributions are observed to narrow slightly for larger $\eta_{\mathrm{gap}}$, reflecting the fact that the true $v_n$ decreases slowly at large $|\eta|$~\cite{Aad:2012bu} and a larger $\eta_{\mathrm{gap}}$ on average selects particles at large $|\eta|$. However, the results are always consistent between the two methods independent of the $\eta_{\mathrm{gap}}$ value used. This consistency supports further the conclusion that the influence of the short-range non-flow correlations on the final unfolded results is not significant. 

The dependence of the EbyE $v_n$ on the $\pT$ of the charged particles has also been checked: the particles are divided into those with $0.5<\pT<1$~GeV and those with $\pT>1$~GeV, and the EbyE $v_n$ measurements are repeated independently for each group of particles. About 60\% of detected particles have $0.5<\pT<1$~GeV, and this fraction varies weakly with centrality. The unfolding performance is found to be slightly worse for charged particles with $0.5<\pT<1$~GeV than for those with $\pT>1$~GeV, due to their much smaller $v_n$ signal. Hence the $v_n$ range of the unfolded distribution for the final results is chosen separately for each $\pT$ range. 

The final $v_n$ distributions are obtained using the single-particle unfolding with the full-ID and $N_{\mathrm{iter}}=128$, separately for charged particles in the two aforementioned $\pT$ ranges and the combined $\pT$ range. The prior is obtained by rescaling the $v_n^{\mathrm{obs}}$ distribution according to eq.~(\ref{eq:pri0}) with $f=0$, and the response function is measured from correlations of the two half-IDs with no $\eta$ gap in between. 

\subsection{Systematic uncertainties}
\label{sec:m2.2}
The systematic uncertainties associated with the unfolding procedure include contributions from the residual non-convergence, dependence on the prior, uncertainty in the response function, the difference between the single-particle method and the 2PC method, and the tracking efficiency. The residual non-convergence is estimated from the difference between the results for $N_{\mathrm{iter}}=32$ and $N_{\mathrm{iter}}=128$ or between the results for the half-ID and full-ID. These two estimates are strongly correlated, so in each bin of the unfolded distribution the larger of the two is used. The dependence on the prior is taken as the difference between the results for $f=0$ and $f=2.5$. Note that the prior for $f=0$ ($f=2.5$) is broader (narrower) than the final unfolded distributions. The uncertainty of the response function is estimated from the difference between results obtained using the analytical formula eq.~(\ref{eq:5b}) and results obtained using the measured distribution, as well as the change in the results when the small dependence of $\delta_{_{\mathrm{2SE}}}$ on the observed $v_n^{\mathrm{obs}}$ is taken into account. Results for the $\pT$-dependence of the $v_n$ distributions (see section~\ref{sec:re} and figure~\ref{fig:perf7}) show that the mean values vary with $\pT$, but, after they are rescaled to a common mean, the resulting shapes are almost identical. Motivated by this finding, every source of systematic uncertainty is decomposed into two components: the uncertainty associated with the $\langle v_n\rangle$ or the $v_n$-scale, and the uncertainty in the shape after adjustment to the same $\langle v_n\rangle$ or the adjusted $v_n$-shape. The uncertainties are then combined separately for the $v_n$-scale and the adjusted $v_n$-shape. Most shape uncertainties can be attributed to $\sigma_{v_n}$, such that the remaining uncertainties on the adjusted $v_n$-shape are generally smaller.

\renewcommand{\tabcolsep}{0.1cm}
\begin{table}[!h]
\centering
\small{\begin{tabular}{|c|cccc|cccc|}\hline 
 &\multicolumn{4}{|c|}{Uncertainty in $\langle v_2\rangle$ or $\sigma_{v_2}$}     & \multicolumn{4}{|c|}{Uncertainty in $\sigma_{v_2}/\langle v_2\rangle$} \tabularnewline\hline
Centrality               &   0--10\%    & 10--30\%   &    30--50\% & 50--70\%     &   0--10\%    & 10--30\%   &    30--50\% & 50--70\%\tabularnewline\hline
Non-convergence [\%]         &   $<$0.1    & $<$0.1    &   $<$0.2   & 3--12       &   0.9        &   0.6      &    0.5--1.4 & 3--11\tabularnewline
Prior           [\%]         &   $<$0.1    & $<$0.1    &   $<$0.2   & 0.2         &   0.6        &   $<$0.3   &     $<$0.2  & 0.2--0.7\tabularnewline
Response function  [\%]      &   0.3--1    & 0.3        &   0.3      & 0.2--1     &   1.0        &   0.7      &     0.7     & 0.6--3\tabularnewline
Compare to 2PC [\%]&   $<$0.2    & $<$0.2    &   $<$0.2    &0.2--7                &  0.5--1.5    &   $<$0.4   &  0.4--0.8   & 1--7\tabularnewline
Efficiency               [\%]&   1.3         & 0.8   &   0.8      & 0.7            &   0.4   &    0.4     &  0.4     & 0.4--0.8\tabularnewline
\multirow{2}{1.3in}{\hfil
Track selection, trigger, stability [\%]}& \multirow{2}{*}{\hfil 2.2}  &  \multirow{2}{*}{\hfil 1.9}   &   \multirow{2}{*}{\hfil 1.7} &      \multirow{2}{*}{\hfil 1.7}   
             &  &&& \tabularnewline& & & & & &&&\tabularnewline
Total experimental[\%]                           &  2.6       &  1.9        &  1.8     &   3.5--14       & 1.6--2.2    &  1.3      & 1--1.8    &3.4--14\tabularnewline\hline
Residual non-flow from~\cite{Jia:2013tja} [\%]      &  1.4--2.3  &  0.7--1.8   &  1.5     &   1.7--3.5      & 0.1--1.5    &  1        & 1         &1.5\tabularnewline\hline\hline
&\multicolumn{4}{|c|}{Uncertainty in $\langle v_3\rangle$ or $\sigma_{v_3}$}       & \multicolumn{4}{|c|}{Uncertainty in $\sigma_{v_3}/\langle v_3\rangle$}\tabularnewline\hline
Centrality              & 0--10\%    & 10--30\%    &    30--50\% & 50--60\%       & 0--10\%    & 10--30\%    &    30--50\% & 50--60\%\tabularnewline\hline
Non-convergence         [\%] &   0.2    & 0.3    &    0.3      &  1.2--5       & 0.2--0.8   &  0.3--0.8   &   0.4--2    & 0.5--4\tabularnewline
Prior                   [\%] &    $<$0.2    & $<$0.2    &   $<$0.2 &0.5--1.4       &  0.6       &  0.2--0.4   &   0.2--1.0  & 3.0\tabularnewline
Response function       [\%] &    0.6     & 0.8       &  0.8--2.4    & 2.9--4.6     &  0.3--0.7  &  0.2--0.5   &   0.9--2.5  & 3--5\tabularnewline
Compare to 2PC [\%]&    0.5     &0.2--0.7        &  0.1        & 0.2         &  0.3--1.6  & 0.4--0.6    &   0.7--2.5  & 1--3\tabularnewline
Efficiency               [\%]&   1.6     & 1.2       & 1            & 0.9          &   $<$0.2   &    $<$0.2   &    $<$0.3   & $<$0.3\tabularnewline
\multirow{2}{1.3in}{\hfil
Track selection, trigger, stability [\%]}& \multirow{2}{*}{\hfil 2.1}    & \multirow{2}{*}{\hfil 1.4}         &  \multirow{2}{*}{\hfil 1.5--2}       &  \multirow{2}{*}{\hfil 2.5--4.5}  & &&&\tabularnewline
& & & & & &&&\tabularnewline
Total experimental[\%]                   &  2.7       & 2.2    &  2--3.3  &  4.2--8.3   &  0.8--2     &     0.6--1.2  &   1.3--4.2   &4.2--7.0\tabularnewline\hline
Residual non-flow from~\cite{Jia:2013tja}[\%]                   &  0.4       & 0.6    &  1.2     &  2.0--2.5   &  0.2        &  0.2   & 0.5         &0.5\tabularnewline\hline\hline
&\multicolumn{4}{|c|}{Uncertainty in $\langle v_4\rangle$ or $\sigma_{v_4}$ } & \multicolumn{4}{|c|}{Uncertainty in $\sigma_{v_4}/\langle v_4\rangle$}\tabularnewline\hline
Centrality                &   0--10\%    & 10--30\% & 30--45\% &     &   0--10\%    & 10--30\% & 30--45\% &  \tabularnewline\hline
Non-convergence          [\%] &  1--2.0      & 1--1.5   & 3.0--5.5    &           &  1--2    & 0.5--1      & 2.0--4.0    & \tabularnewline
Prior                    [\%] &  3.0      & 3.0   & 5.0--7.0  &           &  2.0      &   3.0         & 5.0    &\tabularnewline
Response function        [\%] & 2.5--4.0      & 3.0        & 3.0--5.0    &          &  0.5--2    &  0.6--1.2   &  2.0--2.3   & \tabularnewline
Compare to 2PC[\%] &  0.2-1     &0.3       & 1--4.7    &           &  1--2.5    &   1.2       & 0.5--1.2    &\tabularnewline
Efficiency               [\%] & 2.0          &  1.5     & 1.2    &           &    1         &   0.4       &  $<$0.3    &\tabularnewline
\multirow{2}{1.3in}{\hfil
Track selection, trigger, stability [\%]}&  \multirow{2}{*}{\hfil 3.0}      & \multirow{2}{*}{\hfil 2.7}     &  \multirow{2}{*}{\hfil 3--6}   &   & &&& \tabularnewline
& & & & & &&&\tabularnewline
Total experimental[\%]                   &  5.4         &     5.4      &  8--11      &     &   3.0        &  4.0  & 5--7   &\tabularnewline\hline
Residual non-flow from~\cite{Jia:2013tja}[\%]                   &  0.8--1.4    & 2.9--3.2     &  3--5        &     &  0.2--0.6    &  0.4  & 2.5         &\tabularnewline\hline\hline
\end{tabular}}\normalsize
\caption{Summary of systematic uncertainties as percentages of $\langle v_n\rangle$, $\sigma_{v_n}$ and $\sigma_{v_n}/\langle v_n\rangle$ (\mbox{$n=2$--4}) obtained using charged particles with $\pT>0.5$~GeV. The uncertainties for $\langle v_n\rangle$ and $\sigma_{v_n}$ are similar so the larger of the two is quoted. The uncertainties associated with track selection, the trigger and stability are taken from Ref.~\cite{Aad:2012bu}. For completeness, the model dependent estimates of the residual non-flow effects derived from Ref.~\cite{Jia:2013tja} for unfolding method with the default setup are also listed. Most uncertainties are asymmetric; the quoted numbers refer to the maximum uncertainty range spanned by various centrality intervals in each group.}
\label{tab:sysscal0}
\end{table}

To estimate the uncertainty due to the tracking efficiency, the measurement is repeated without applying the efficiency re-weighting. The final distributions are found to have almost identical shape, while the values of $\langle v_n\rangle$ and $\sigma_{v_n}$ increase by a few percent. This increase can be ascribed mainly to the smaller fraction of low-$\pT$ particles, which have smaller $v_n$, so this increase should not be considered as a systematic uncertainty on the $v_n$-scale. Instead, the scale uncertainty is more appropriately estimated from the change in the $v_n^{\mathrm{EP}}$ when varying the efficiency correction within its uncertainty range. On the other hand, small changes are observed for $\sigma_{v_n}/\langle v_n\rangle$ and the adjusted $v_n$-shape. Since these changes are small, they are conservatively included in the total systematic uncertainty in the $v_n$-shape.

The variation of the efficiency with the detector occupancy may reduce the $v_n$ coefficients in eq.~\ref{eq:flow}. This influence has been studied by comparing the $v_n$ values reconstructed via the two-particle correlation method with the generated $v_n$ signal in HJING simulation with flow imposed on the generated particles. The influence is found to be 1\% or less, consistent with the findings in~\cite{ATLAS:2011ah}, and it is included in the uncertainty due to tracking efficiency. It should be pointed out that the influence of detector occupancy is expected to be proportional to the magnitude of the $v_n$ signal, and hence it mainly affects the $v_n$ scale, not the adjusted $v_n$ shape.

Additional systematic uncertainties include those from the track selection, dependence on the running period, and trigger and event selections. These account for the influence of fake tracks, the instability of the detector acceptance and efficiency, variation of the centrality definition, respectively. All three sources of systematic uncertainty are expected to change only the $v_n$-scale but not the $v_n$-shape, and they are taken directly from the published $v_n^{\mathrm{EP}}$ measurement~\cite{Aad:2012bu}. 

Table~\ref{tab:sysscal0} summarizes the various systematic uncertainties as percentages of $\langle v_n\rangle$, $\sigma_{v_n}$ and $\sigma_{v_n}/\langle v_n\rangle$, for charged particles with $\pT>0.5$~GeV. The uncertainties on $\langle v_n\rangle$ and $\sigma_{v_n}$ are strongly correlated, which in many cases leads to a smaller and asymmetric uncertainty on $\sigma_{v_n}/\langle v_n\rangle$. In most cases, the uncertainties are dominated by tracking efficiency, as well as the track selection, trigger, and run stability assessed in Ref.~\cite{Aad:2012bu}. The uncertainties associated with the unfolding procedure are usually significant only in peripheral collisions, except for $\langle v_4\rangle$, where they are important across the full centrality range. The relative uncertainties in table~\ref{tab:sysscal0} have also been evaluated separately for charged particles with $0.5<\pT<1$~GeV and $\pT>1$~GeV. In general, the systematic uncertainties are larger for the group of particles with $0.5<\pT<1$~GeV, due mainly to the increased contributions from residual non-convergence and the choice of priors.

The final $v_n$ distributions are shown over a $v_n$ range that is chosen such that the statistical uncertainties in all bins are less than 15\%, and the results obtained with the default setups between $N_{\mathrm{iter}}=32$ and $N_{\mathrm{iter}}=128$ are consistent within 10\%. The systematic uncertainties on the adjusted shape from the sources discussed above are then combined to give the final uncertainty for the $v_n$-shape. The total systematic uncertainties are typically a few percent of $v_n$ in the main region of the $v_n$ distributions and increase to 15\%-30\% in the tails, depending on the value of $n$ and centrality interval (see figure~\ref{fig:result1}). Within the chosen $v_n$ ranges, the statistical uncertainties are found to be always smaller than the systematic uncertainties for the $v_n$-shape, and the integrals of the $v_n$ distributions outside these ranges are typically $<0.5\%$ for the 5\%-wide centrality intervals and $<1\%$ for the 1\%-wide centrality intervals.

This analysis relies on the data-driven unfolding method to suppress the non-flow effects. In fact, many of the cross-checks presented in section~\ref{sec:m2.1} are sensitive to the residual non-flow, where ``residual non-flow'' refers to that component of the non-flow effects that is not removed by the unfolding method. Hence the total experimental systematic uncertainties quoted for the $v_n$-scale in table~\ref{tab:sysscal0} and the systematic uncertainties on the $v_n$-shape discussed above already include an estimate of the residual non-flow effects. An alternative, albeit model-dependent approach, is to rely on simulations. One such study is carried out in Ref.~\cite{Jia:2013tja} based on HIJING with EbyE $v_n$ imposed on the generated particles. This study demonstrates that most non-flow effects are indeed suppressed by the data-driven unfolding method used in this analysis. This study also shows that the residual non-flow effects for the unfolding method with the default setup have no appreciable impact on the $v_3$ distributions, but broaden slightly the $v_2$ and $v_4$ distributions. Furthermore, Ref.~\cite{Jia:2013tja} also shows that most of these changes can be absorbed into simultaneous increases of $\langle v_n\rangle$ and $\sigma_{v_n}$ values by a few percent. For completeness, these model-dependent estimates of the residual non-flow contribution on the $v_n$-scale are quoted in table~\ref{tab:sysscal0}: they are found to be smaller than the total systematic uncertainties of the measurement, especially for the higher-order harmonics. Ref.~\cite{Jia:2013tja} also estimates the influence of residual non-flow on the intrinsic $v_n$-shape, obtained by comparing the shapes of the unfolded and the input distributions after both distributions are rescaled to have the same $\langle v_n\rangle$. The variations of the intrinsic $v_n$-shape due to residual non-flow are found to reach a maximum of 5\%--15\% (of $p(v_n)$) in the tails of the distributions, depending on the choice of $n$ and the centrality interval, but these variations are typically much smaller than the total systematic uncertainties on the $v_n$-shape in the data (see figure~\ref{fig:result1}). As noted above,  a large fraction of the residual non-flow effects estimated by Ref.~\cite{Jia:2013tja} are expected to be already included in various data-driven cross-checks performed in section~\ref{sec:m2.1}. Hence, to avoid double counting, the total systematic uncertainties applied to the data plots do not include the estimates from Ref.~\cite{Jia:2013tja}. However, the uncertainties from Ref.~\cite{Jia:2013tja} for the $v_n$-scale (table~\ref{tab:sysscal0}) and $v_n$-shape, are well within the total systematic uncertainties derived from the data analysis.

\section{Results}
\label{sec:re}
\begin{figure}[!t]
\centering
\includegraphics[width=1\columnwidth]{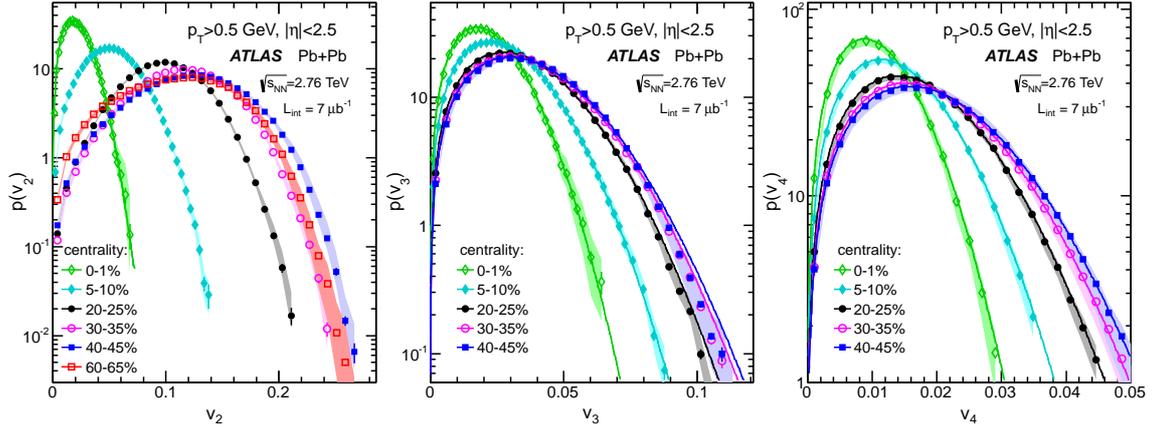}\vspace*{-0.2cm}
\caption{\label{fig:result1} The probability density distributions of the EbyE $v_n$ in several centrality intervals for $n=2$ (left panel), $n=3$ (middle panel) and $n=4$ (right panel). The error bars are statistical uncertainties, and the shaded bands are uncertainties on the $v_n$-shape. The solid curves are distributions calculated from the measured $\langle v_n\rangle$ according to eq.~(\ref{eq:fluc2b}). The solid curve is shown only for 0--1\% centrality interval for $v_2$, but for all centrality intervals in case of $v_3$ and $v_4$.}
\end{figure}

Figure~\ref{fig:result1} shows the probability density distributions of the EbyE $v_n$ in several centrality intervals obtained for charged particles with $\pT>0.5$~GeV. The shaded bands indicate the systematic uncertainties associated with the shape. These uncertainties are strongly correlated in $v_n$: the data points are allowed to change the shape of the distribution within the band while keeping $\langle v_n\rangle$ unchanged. The $v_n$ distributions are found to broaden from central to peripheral collisions (especially for $v_2$), reflecting the gradual increase of the magnitude of $v_n$ for more peripheral collisions~\cite{ATLAS:2011ah,Aad:2012bu}. The shape of these distributions changes quickly with centrality for $v_2$, while it changes more slowly for higher-order harmonics. These distributions are compared with the probability density function obtained from eq.~(\ref{eq:fluc2b}) ($v_n^{\mathrm{\;RP}}=0$), which represents a fluctuation-only scenario for $v_n$. These functions, indicated by the solid curves, are calculated directly from the measured $\langle v_n\rangle$ values via eq.~(\ref{eq:fluc3}) for each distribution. The fluctuation-only scenario works reasonably well for $v_3$ and $v_4$ over the measured centrality range, but fails for $v_2$ except for the most central 2\% of collisions, i.e. for the centrality interval 0-2\%. Hence for $v_2$ the solid curve representing the fluctuation-only scenario is shown only for the 0-1\% centrality interval (the data for the 1-2\% interval are not shown). However, there is a small systematic difference between the data and the curve in the tails of the $v_3$ distributions in mid-central collisions, with a maximum difference of two standard deviations. Using eq.~(\ref{eq:fluc2c}), this difference is compatible with a non-zero $v_3^{\mathrm{\;RP}}$ similar to the findings reported in Ref.~\cite{ALICE:2011ab}. Futhermore, since the measured $v_4$ distribution covers only a limited range ($v_4\lesssim3\delta_{_{v_4}}$), a non-zero $v_4^{\mathrm{\;RP}}$ on the order of $\delta_{_{v_4}}$ can not be excluded by this analysis based on eq.~(\ref{eq:fluc2c}).

\begin{figure}[!t]
\centering
\includegraphics[width=1\columnwidth]{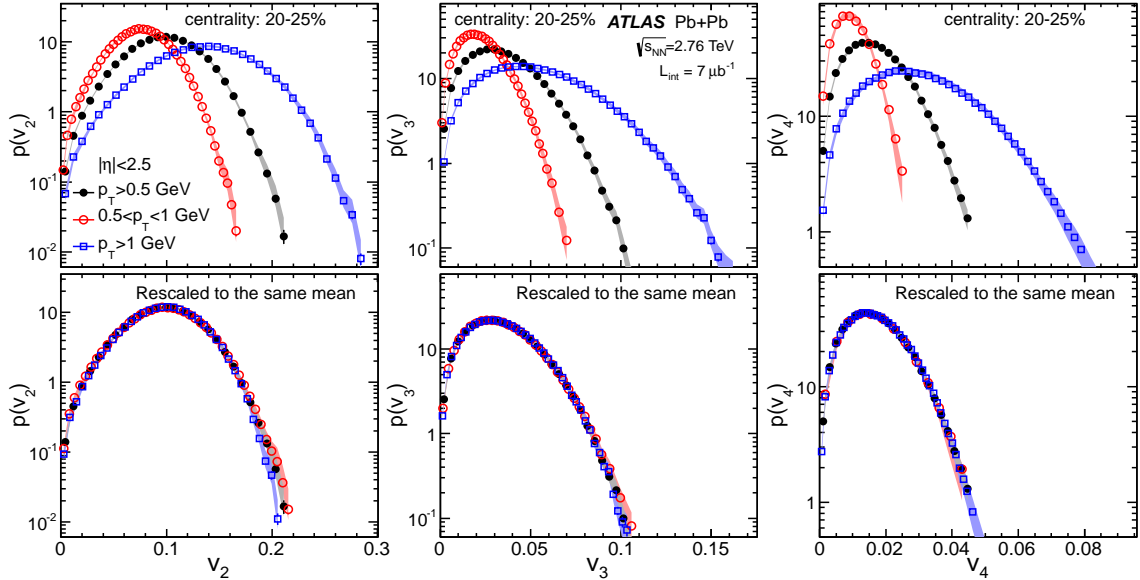}\vspace*{-0.2cm}
\caption{\label{fig:perf7}  Top panels: The unfolded distributions for $v_n$ in the 20--25\% centrality interval for charged particles in the $\pT>0.5$~GeV, $0.5<\pT<1$~GeV and $\pT>1$~GeV ranges. Bottom panels: same distributions but rescaled horizontally so the $\langle v_n\rangle$ values match that for the $\pT>0.5$~GeV range. The shaded bands represent the systematic uncertainties on the $v_n$-shape.}
\end{figure}
Figure~\ref{fig:perf7} compares the unfolded $v_n$ distributions for charged particles in three $\pT$ ranges: $\pT>0.5$~GeV, $0.5<\pT<1$~GeV and $\pT>1$~GeV. The $v_n$ distributions for $\pT>1$~GeV are much wider than for $0.5<\pT<1$~GeV, reflecting the fact that the $v_n$ values increase strongly with $\pT$ in this region~\cite{Aad:2012bu}. However, once these distributions are rescaled to the same $\langle v_n\rangle$ as shown in the lower row of figure~\ref{fig:perf7}, their shapes are quite similar except in the tails of the distributions for $n=2$. This behaviour suggests that the hydrodynamic response of the medium to fluctuations in the initial geometry is nearly independent of $\pT$ in the low-$\pT$ region; it also demonstrates the robustness of the unfolding performance against the change in the underlying $v_n$ distributions and response functions.

\begin{figure}[!t]
\centering
\includegraphics[width=1\columnwidth]{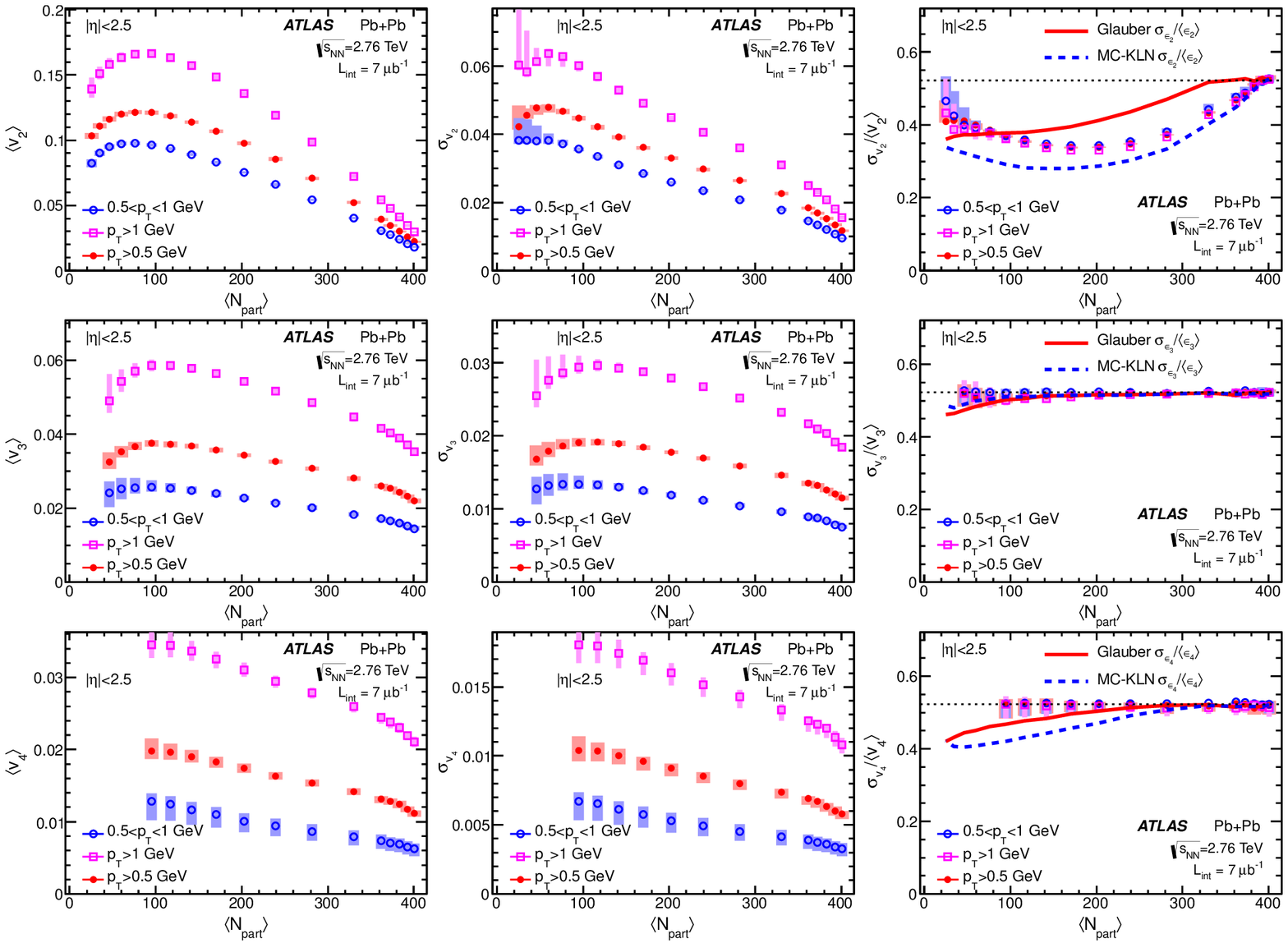}\vspace*{-0.2cm}
\caption{\label{fig:result2}  The $\langle N_{\mathrm{part}}\rangle$ dependence of $\langle v_n\rangle$ (left panels), $\sigma_{v_n}$ (middle panels) and $\sigma_{v_n}/\langle v_n\rangle$ (right panels) for $n=2$ (top row), $n=3$ (middle row) and $n=4$ (bottom row). Each panel shows the results for three $\pT$ ranges together with the total systematic uncertainties. The dotted lines in the right column indicate the value $\sqrt{4/\pi-1}$ expected for the radial projection of a 2D Gaussian distribution centred around origin (see eq.~(\ref{eq:large})). The values of $\sigma_{v_n}/\langle v_n\rangle$ are compared with the $\sigma_{\epsilon_n}/\langle \epsilon_n\rangle$ given by the Glauber model~\cite{Miller:2007ri} and MC-KLN model~\cite{Drescher:2006pi}.
}
\end{figure}

Figure~\ref{fig:result2} shows a summary of the quantities derived from the EbyE $v_n$ distributions, i.e. $\langle v_n\rangle$, $\sigma_{v_n}$ and $\sigma_{v_n}/\langle v_n\rangle$, as a function of $\langle N_{\mathrm{part}}\rangle$. The shaded bands represent the total systematic uncertainties listed in table~\ref{tab:sysscal0}, which generally are asymmetric. 
Despite the strong $\pT$ dependence of $\langle v_n\rangle$ and $\sigma_{v_n}$, the ratio $\sigma_{v_n}/\langle v_n\rangle$ is relatively stable. For $v_2$, the value of $\sigma_{v_n}/\langle v_n\rangle$ varies strongly with $\langle N_{\mathrm{part}}\rangle$, and reaches a minimum of about 0.34 at $\langle N_{\mathrm{part}}\rangle\sim200$, corresponding to the 20--30\% centrality interval. For $v_3$ and $v_4$, the values of $\sigma_{v_n}/\langle v_n\rangle$ are almost independent of $\langle N_{\mathrm{part}}\rangle$, and are consistent with the value expected from the fluctuation-only scenario ($\sqrt{4/\pi-1}$ via eq.~(\ref{eq:large}) as indicated by the dotted lines), except for a small deviation for $v_3$ in mid-central collisions. This limit is also reached for $v_2$ in the most central collisions as shown by the top-right panel of figure~\ref{fig:result2}. 

\begin{figure}[!t]
\centering
\includegraphics[width=1\columnwidth]{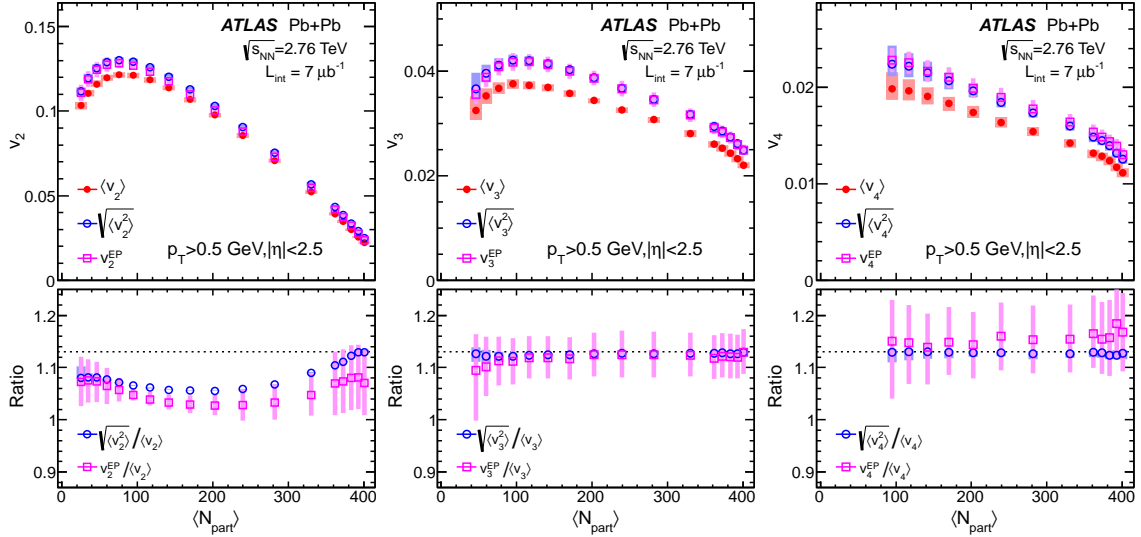}
\caption{\label{fig:result3a} Top panels: Comparison of $\langle v_n\rangle$ and $\sqrt{\langle v_n^2\rangle}\equiv\sqrt{\langle v_n\rangle^2+\sigma_{v_n}^2}$, derived from the EbyE $v_n$ distributions, with the $v_n^{\mathrm{EP}}$~\cite{Aad:2012bu}. Bottom panels: the ratios of $\sqrt{\langle v_n^2\rangle}$ and  $v_n^{\mathrm{EP}}$ to $\langle v_n\rangle$. The shaded bands represent the systematic uncertainties. The dotted lines in bottom panels indicate $\sqrt{\langle v_n^2\rangle}/\langle v_n\rangle=1.13$, expected for the radial projection of a 2D Gaussian distribution centred around origin (eq.~(\ref{eq:large})).}
\end{figure}

Figure~\ref{fig:result3a} compares the $\langle v_n\rangle$ and $\sqrt{\langle v_n^2\rangle}$ with the $v_n^{\mathrm{EP}}$ measured using the FCal event plane method for charged particles with $\pT>0.5$~GeV~\cite{Aad:2012bu}. For $v_3$ and $v_4$, the values of $v_n^{\mathrm{EP}}$ are almost identical to $\sqrt{\langle v_n^2\rangle}$. However, the values of $v_2^{\mathrm{EP}}$ are in between $\langle v_2\rangle$ and $\sqrt{\langle v_2^2\rangle}$. As expected~\cite{Alver:2008zza,Ollitrault:2009ie}, and as discussed in section~\ref{sec:m1.3}, they approach $\sqrt{\langle v_2^2\rangle}$ only in peripheral collisions, where the resolution factor used in the EP method is small.

\begin{figure}[!h]
\centering
\includegraphics[width=1\columnwidth]{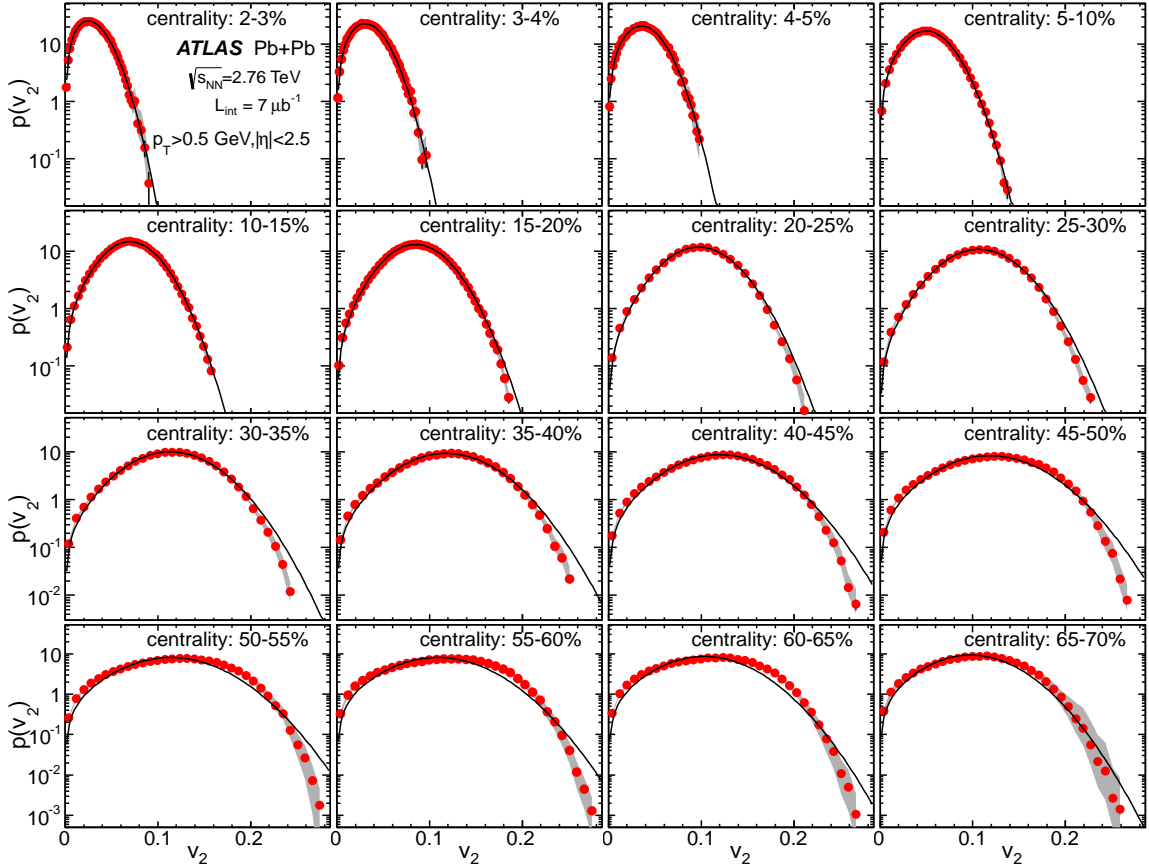}
\caption{\label{fig:fit} The probability density distributions of $v_2$ for $\pT>0.5$ GeV in several centrality intervals, together with fits to the Bessel--Gaussian function eq.~(\ref{eq:fluc2}). The fits for the 0--1\% and 1--2\% centrality intervals are not shown but they can be well described by a Bessel--Gaussian function with $v_2^{\mathrm{RP}}=0$ (see discussion for figure~\ref{fig:result1}). The $\langle N_{\mathrm{part}}\rangle$ value for each centrality range is given in table~\ref{tab:cent}.}
\end{figure}

\begin{figure}[!h]
\centering
\includegraphics[width=1\columnwidth]{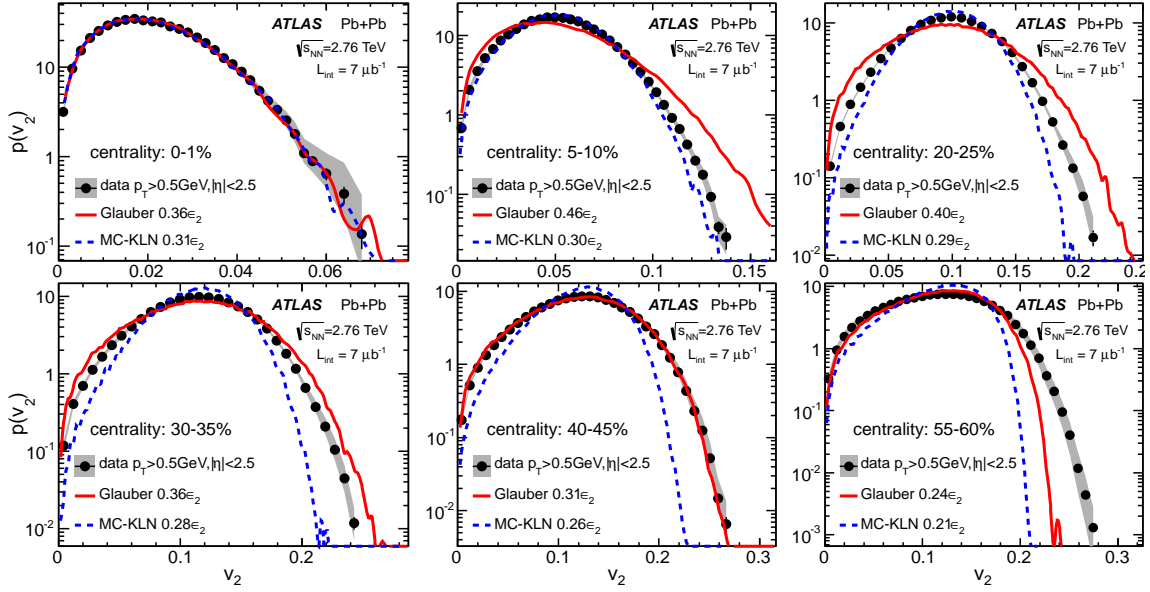}
\caption{\label{fig:result4a} The EbyE $v_2$ distributions compared with the $\epsilon_2$ distributions from two initial geometry models: a Glauber model (solid lines) and the MC-KLN model (dashed lines). The $\epsilon_2$ distributions have been rescaled to the same mean values. The scale factors are indicated in the legends.}
\end{figure}

The results in figures~\ref{fig:result1} and \ref{fig:result2} imply that the distributions of $v_2$ in central collisions (centrality interval 0-2\%), and of $v_3$ and $v_4$ in most of the measured centrality range are described by a 2D Gaussian function of $\overrightharp{v}_n$ centred around the origin (eq.~\ref{eq:fluc2b}). On the other hand, the deviation of the $v_2$ distribution from such a description in other centrality intervals suggests that the contribution associated with the average geometry, $v_2^{\mathrm{RP}}$, becomes important. In order to test this hypothesis, the $v_2$ distributions have been fitted to the Bessel-Gaussian function eq.~(\ref{eq:fluc2}), with $v_2^{\mathrm{RP}}$ not constrained to be zero. The results of the fit are shown in figure~\ref{fig:fit} for various centrality intervals. The fit works reasonably well up to the 25--30\% centrality interval, although systematic deviations in the tails are apparent already in the 15--20\% centrality interval. The deviations increase steadily for more peripheral collisions, which may be due to the fact that the fluctuations of $\epsilon_2$ (eq.~\ref{eq:ena}) are no longer Gaussian in peripheral collisions where $N_{\mathrm{part}}$ is small~\cite{Alver:2008zza} (see also figure~\ref{fig:result4a}).

The values of $v_2^{\mathrm{RP}}$ and $\delta_{_{v_2}}$ can be estimated from these fits. Since the value of $v_2^{\mathrm{RP}}$ varies rapidly with $\langle N_{\mathrm{part}}\rangle$, especially in central collisions, the extracted $v_2^{\mathrm{RP}}$ and $\delta_{_{v_2}}$ values can be affected by their spreads if too broad a centrality interval is used. The effect of the centrality binning has been checked and corrected as follows. Taking the 20--25\% interval as an example, the results obtained using the full centrality range within this interval are compared to the results obtained from the average of the five individual 1\% intervals: 20--21\%,...,24--25\%. This procedure has been carried out for each 5\% centrality interval, and the difference is found to be significant only for the 0--5\% and 5--10\% intervals, and negligible for all the others. For the 0--5\% interval, results are reported in the individual 1\% bins. For the 5--10\% bin, results are averaged over the five individual 1\% bins.

As a cross-check, the Bessel-Gaussian fits are also performed on the $v_2^{\mathrm{obs}}$ distributions before the unfolding. Systematic devations are also observed between the fit and the $v_2^{\mathrm{obs}}$ data, but the deviations are smaller than those shown in figure~\ref{fig:fit}. The value of $v_2^{\mathrm{RP}}$ from the $v_2^{\mathrm{obs}}$ distribution is found to agree to within a few percent with that from the unfolded $v_2$ distribution, while the value of $\delta_{_{v_2}}$ from the $v_2^{\mathrm{obs}}$ distribution is significantly larger. This behaviour is expected since the smearing by the response function (eq.~\ref{eq:5b}) increases mainly the width, and the value of $v_2^{\mathrm{RP}}$ should be stable.

\begin{figure}[!h]
\centering
\includegraphics[width=1\columnwidth]{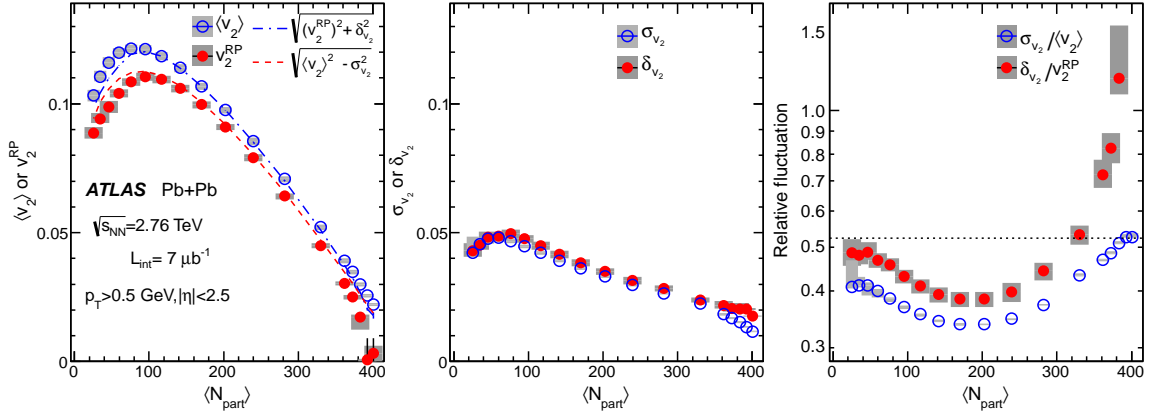}
\caption{\label{fig:fit2} The dependence of $v_2^{\mathrm{RP}}$ and $\langle v_2\rangle$ (left panel),  $\delta_{_{v_2}}$ and $\sigma_{v_2}$ (middle panel), $\delta_{_{v_2}}/v_2^{\mathrm{RP}}$ and $\sigma_{_{v_2}}/\langle v_2\rangle$ (right panel) on $\langle N_{\mathrm{part}}\rangle$. The shaded boxes indicate the systematic uncertainties. The dotted line in the right panel indicates the value $\sqrt{4/\pi-1}$ expected for the radial projection of a 2D Gaussian distribution centred around origin (see eq.~(\ref{eq:large})).}
\end{figure}

Figure~\ref{fig:fit2} shows the $v_2^{\mathrm{RP}}$ and $\delta_{_{v_2}}$ values extracted from the $v_2$ distributions as a function of $\langle N_{\mathrm{part}}\rangle$. They are compared with values of $\langle v_2\rangle$ and $\sigma_{v_2}$ obtained directly from the $v_2$ distributions. The $v_2^{\mathrm{RP}}$ value is always smaller than the value for $\langle v_2\rangle$, and it decreases to zero in the 0--2\% centrality interval, consistent with the results shown in figure~\ref{fig:result1}. The value of $\delta_{_{v_2}}$ is close to $\sigma_{_{v_2}}$ except in the most central collisions. This behaviour leads to a value of $\delta_{_{v_2}}/v_2^{\mathrm{RP}}$ larger than $\sigma_{_{v_2}}/\langle v_2\rangle$ over the full centrality range as shown in the right panel of figure~\ref{fig:fit2}. The value of $\delta_{_{v_2}}/v_2^{\mathrm{RP}}$ decreases with $\langle N_{\mathrm{part}}\rangle$ and reaches a minimum of $0.38\pm0.02$ at $\langle N_{\mathrm{part}}\rangle\approx200$, but then increases for more central collisions. The two points for the 0--1\% and 1--2\% centrality intervals are omitted as the corresponding $v_2^{\mathrm{RP}}$ values are consistent with zero. 

According to eq.~(\ref{eq:small}), when the relative fluctuations are small the value of $\langle v_n\rangle$ can be approximated by:
\begin{eqnarray}
\label{eq:small1}
\langle v_n\rangle\approx \sqrt{ \left(v_n^{\mathrm{RP}}\right)^2+\delta^2_{_{v_n}}}\;.
\end{eqnarray}
Similarly, the value of $v_n^{\mathrm{RP}}$ can be estimated from $\sigma_{_{v_n}}$ and $\langle v_n\rangle$ without relying on the fit:
\begin{eqnarray}
\label{eq:cumu}
 v_n^{\mathrm{RP}}\approx \sqrt{\langle v_n\rangle^2-\sigma_{v_n}^2}\;.
\end{eqnarray}
Both relations are shown in the left panel of figure~\ref{fig:fit2} for $v_2$. Good agreement with the data is observed for $100<\left\langle N_{\mathrm{part}}\right\rangle<350$, corresponding to 5--45\% centrality interval in table~\ref{tab:cent}. However, systematic deviations are observed both in central collisions where fluctuations are dominant, and in peripheral collisions where the Bessel--Gaussian function fails to describe the shape of the $v_2$ distribution.

Multi-particle cumulant methods have been widely used to estimate the values of $v_2^{\mathrm{RP}}$ and $\delta_{_{v_2}}$, as well as to study the deviation of the $v_2$ distribution from a Bessel--Gaussian description~\cite{Agakishiev:2011eq}. The second-order coefficients for the 2$k^{\mathrm{th}}$-order cumulants, $v_{2}\{2k\}$, are generally calculated via $2k$-particle correlations~\cite{Borghini:2000sa}. These coefficients can also be calculated analytically from the measured $v_2$ distributions, labelled as $v_{2}^{\mathrm{calc}}\{2k\}$. The first four terms can be expressed~\cite{Voloshin:2007pc} as:
\begin{eqnarray}
\nonumber
v_{2}^{\mathrm{calc}}\{2\}^2&\equiv& \langle v_2^2\rangle\approx \left(v_2^{\mathrm{RP}}\right)^2+2\delta^2_{_{v_2}}\;,\\\nonumber
v_{2}^{\mathrm{calc}}\{4\}^4&\equiv& -\langle v_2^4\rangle+2\langle v_2^2\rangle^2\approx \left(v_2^{\mathrm{RP}}\right)^4\;,\\\nonumber
v_{2}^{\mathrm{calc}}\{6\}^6&\equiv& \left(\langle v_2^6\rangle^2-9\langle v_2^4\rangle\langle v_2^2\rangle+12\langle v_2^2\rangle^3\right)/4\approx \left(v_2^{\mathrm{RP}}\right)^6\;,\\\nonumber
v_{2}^{\mathrm{calc}}\{8\}^8&\equiv& -\left(\langle v_2^8\rangle^2-16\langle v_2^6\rangle\langle v_2^2\rangle-18\langle v_2^4\rangle^2+144\langle v_2^4\rangle\langle v_2^2\rangle^2-144\langle v_2^2\rangle^4\right)/33\approx \left(v_2^{\mathrm{RP}}\right)^8\;.\\\label{eq:cumu2}
\end{eqnarray}
The last part of each equation is exact when the $v_2$ distribution follows the Bessel--Gaussian function eq.~(\ref{eq:fluc2}). In this case only the first two cumulants are independent, and the higher-order cumulants do not provide new information. For the same reason, it has been argued that differences between the higher-order cumulants $v_{2}\{2k\}$ ($k>1$) are sensitive to non-Gaussian behaviour in the underlying $v_2$ distribution~\cite{Voloshin:2007pc}.
\begin{figure}[!h]
\centering
\includegraphics[width=0.6\columnwidth]{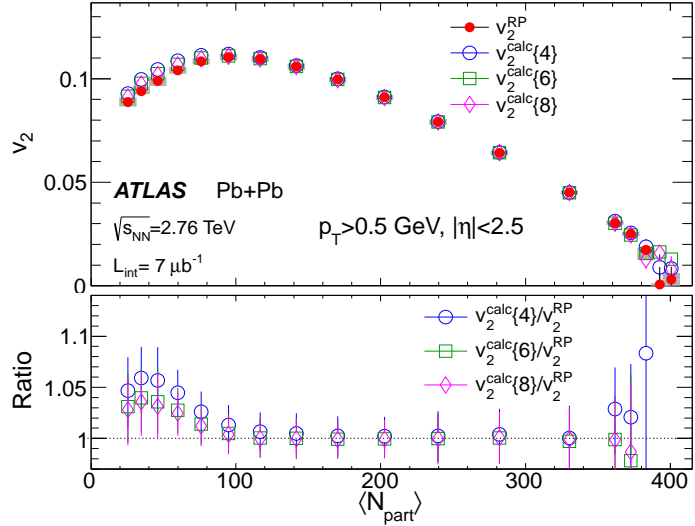}
\caption{\label{fig:fit3} Top panel: Comparison of the $v_2^{\mathrm{RP}}$ obtained from the Bessel--Gaussian fit of the $v_2$ distributions with the values for four-particle ($v_2^{\mathrm{calc}}\{4\}$), six-particle ($v_2^{\mathrm{calc}}\{6\}$) and eight-particle ($v_2^{\mathrm{calc}}\{8\}$) cumulants calculated directly from the unfolded $v_2$ distributions, using eq.~(\ref{eq:cumu2}). Bottom panel: The ratios of the cumulants to the fit results, with the error bars representing the total uncertainties.}
\end{figure}

Figure~\ref{fig:fit3} shows results for $v_{2}^{\mathrm{calc}}\{4\}$,  $v_{2}^{\mathrm{calc}}\{6\}$ and $v_{2}^{\mathrm{calc}}\{8\}$, calculated directly from the measured $v_2$ distributions (eq.~(\ref{eq:cumu2})). They are compared with the results for $v_2^{\mathrm{RP}}$ obtained from the Bessel-Gaussian fits shown in figure~\ref{fig:fit}. The results calculated from the cumulants agree with the results obtained from the fit, except for $\langle N_{\mathrm{part}}\rangle<100$. However, the calculated coefficients from higher-order cumulants agree with each other over the whole centrality range to within 0.5\%-2\%, despite the poor description of the $v_2$ distribution by the Bessel--Gaussian function for the centrality interval 25--70\% ($\langle N_{\mathrm{part}}\rangle<200$), shown in figure~\ref{fig:fit}. It therefore follows that similar values of $v_{2}\{2k\}$ for $k\geq2$ observed in previous measurements~\cite{Agakishiev:2011eq,Abelev:2012di} are no guarantee that the distribution is well described by a Bessel--Gaussian over the full range in non-central collisions, since the uncertainties of these measurements are bigger than the differences seen between $v_{2}^{\mathrm{calc}}\{4\}$, $v_{2}^{\mathrm{calc}}\{6\}$ and $v_{2}^{\mathrm{calc}}\{8\}$ in figure~\ref{fig:fit3}.

\begin{figure}[!h]
\centering
\includegraphics[width=1\columnwidth]{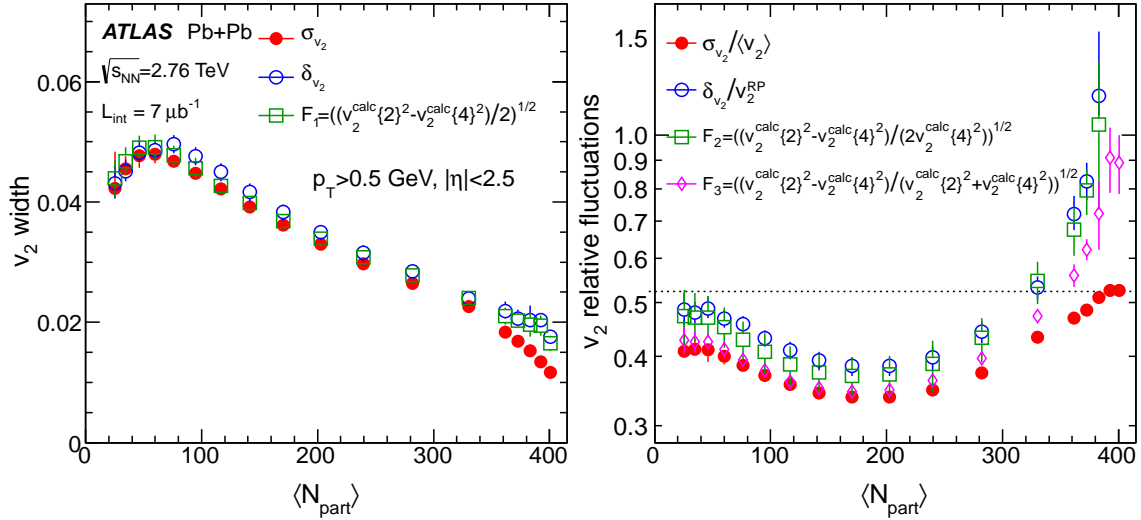}
\caption{\label{fig:compmethod} Left panel: The standard deviation ($\sigma_{_{v_2}}$), the width obtained from Bessel--Gaussian function ($\delta_{_{v_2}}$), the width estimated from the two-particle cumulant ($v_2^{\mathrm{calc}}\{2\}$) and four-particle cumulant ($v_2^{\mathrm{calc}}\{4\}$). These cumulants are calculated analytically via eq.~(\ref{eq:cumu2}) from the $v_2$ distribution. Right panel: Various estimates of the relative fluctuations (see legend).}
\end{figure}

In previous analyses based on cumulants~\cite{Voloshin:2008dg,Abelev:2012di}, several other quantities have been used to study the $v_2$ fluctuations. The definitions of these quantities are given below and their physical meaning can be understood from eq.~(\ref{eq:cumu2}): 
\begin{eqnarray}
\label{eq:cumu3}
F_1&=&{((v_2^{\mathrm{calc}}\{2\}^2-v_2^{\mathrm{calc}}\{4\}^2)/2)^{1/2}\approx \delta_{_{v_2}}},\\
F_2&=&{((v_2^{\mathrm{calc}}\{2\}^2-v_2^{\mathrm{calc}}\{4\}^2)/(2v_2^{\mathrm{calc}}\{4\}^2))^{1/2}}\approx \delta_{_{v_2}}/v_2^{\mathrm{RP}}\\\label{eq:cumu4}
F_3&=&((v_2^{\mathrm{calc}}\{2\}^2-v_2^{\mathrm{calc}}\{4\}^2)/(v_2^{\mathrm{calc}}\{2\}^2+v_2^{\mathrm{calc}}\{4\}^2))^{1/2}\approx \delta_{_{v_2}}/\langle v_2\rangle,
\end{eqnarray}
where $F_1$, $F_2$, and $F_3$ are calculated from the unfolded distributions, using eq.~(\ref{eq:cumu2}). The approximation for $F_3$ is valid when $v_2^{\mathrm{RP}}\gg\delta_{_{v_2}}$. In central collisions where $v_2^{\mathrm{RP}}\muchless \delta_{_{v_2}}$, the value of $F_3$ is expected to approach one. 

Figure~\ref{fig:compmethod} compares the calculated values of $F_1$, $F_2$ and $F_3$ to the rightmost expressions in eqs.~(\ref{eq:cumu3})-(\ref{eq:cumu4}), using $\delta_{v_2}$, $v_2^{\mathrm{RP}}$ obtained from fits to the Bessel-Gaussian function, and the mean of the unfolded distribution. The value of $F_1$ is between $\sigma_{_{v_2}}$ and $\delta_{_{v_2}}$. The quantities $F_2$ and $F_3$ show similar $\langle N_{\mathrm{part}}\rangle$ dependence as $\sigma_{_{v_2}}/\langle v_2\rangle$ and $\delta_{_{v_2}}/v_2^{\mathrm{RP}}$, however significant discrepancies are observed, especially in the most central collisions where the flow fluctuation is dominant.

\begin{figure}[!h]
\centering
\includegraphics[width=1\columnwidth]{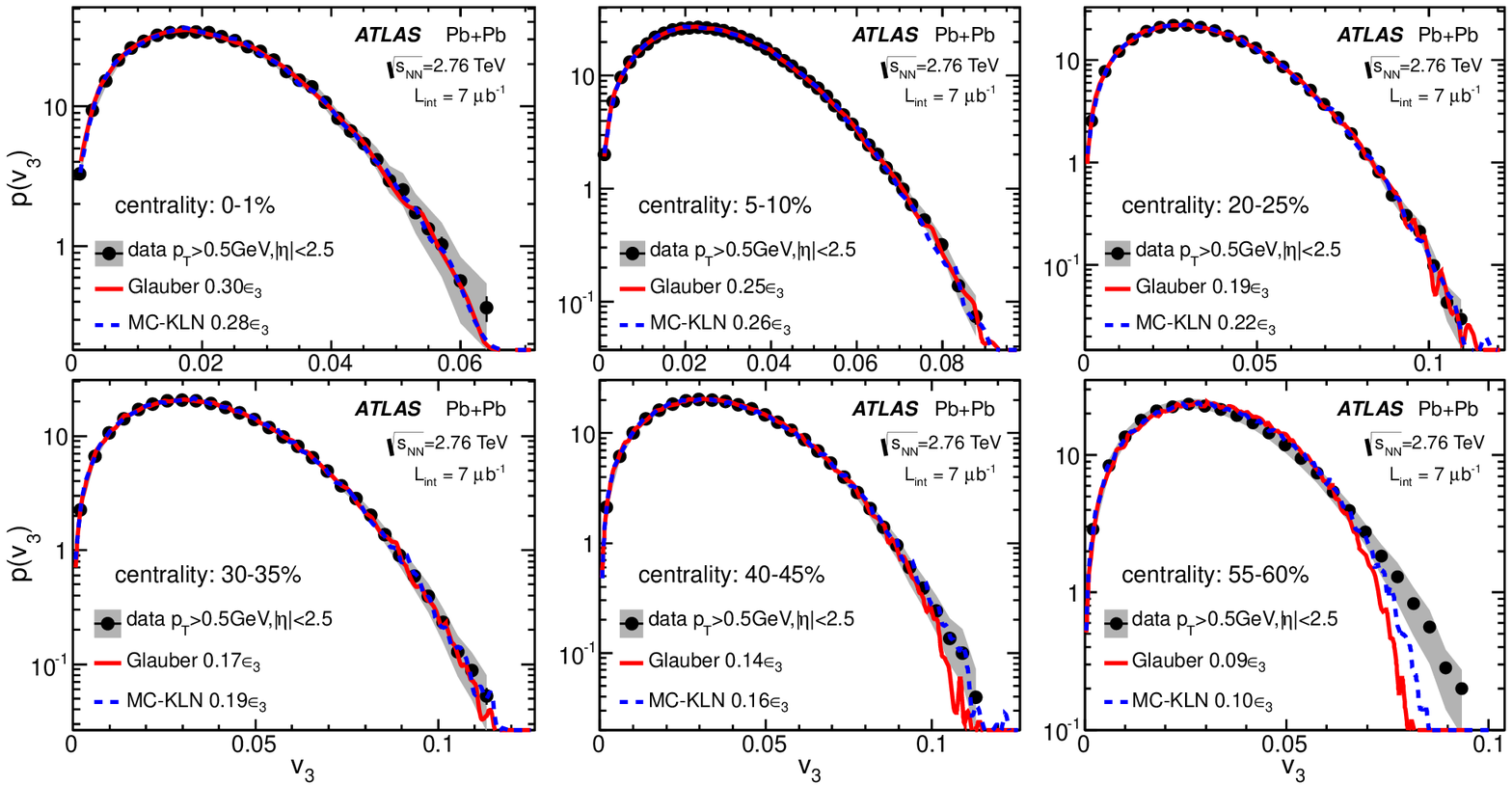}
\caption{\label{fig:result4b} The EbyE $v_3$ distributions compared with the $\epsilon_3$ distributions from two initial geometry models: a Glauber model (solid lines) and the MC-KLN model (dashed lines). The $\epsilon_3$ distributions have been rescaled to the same mean values. The scale factors are indicated in the legends.}
\end{figure}
\begin{figure}[!h]
\centering
\includegraphics[width=1\columnwidth]{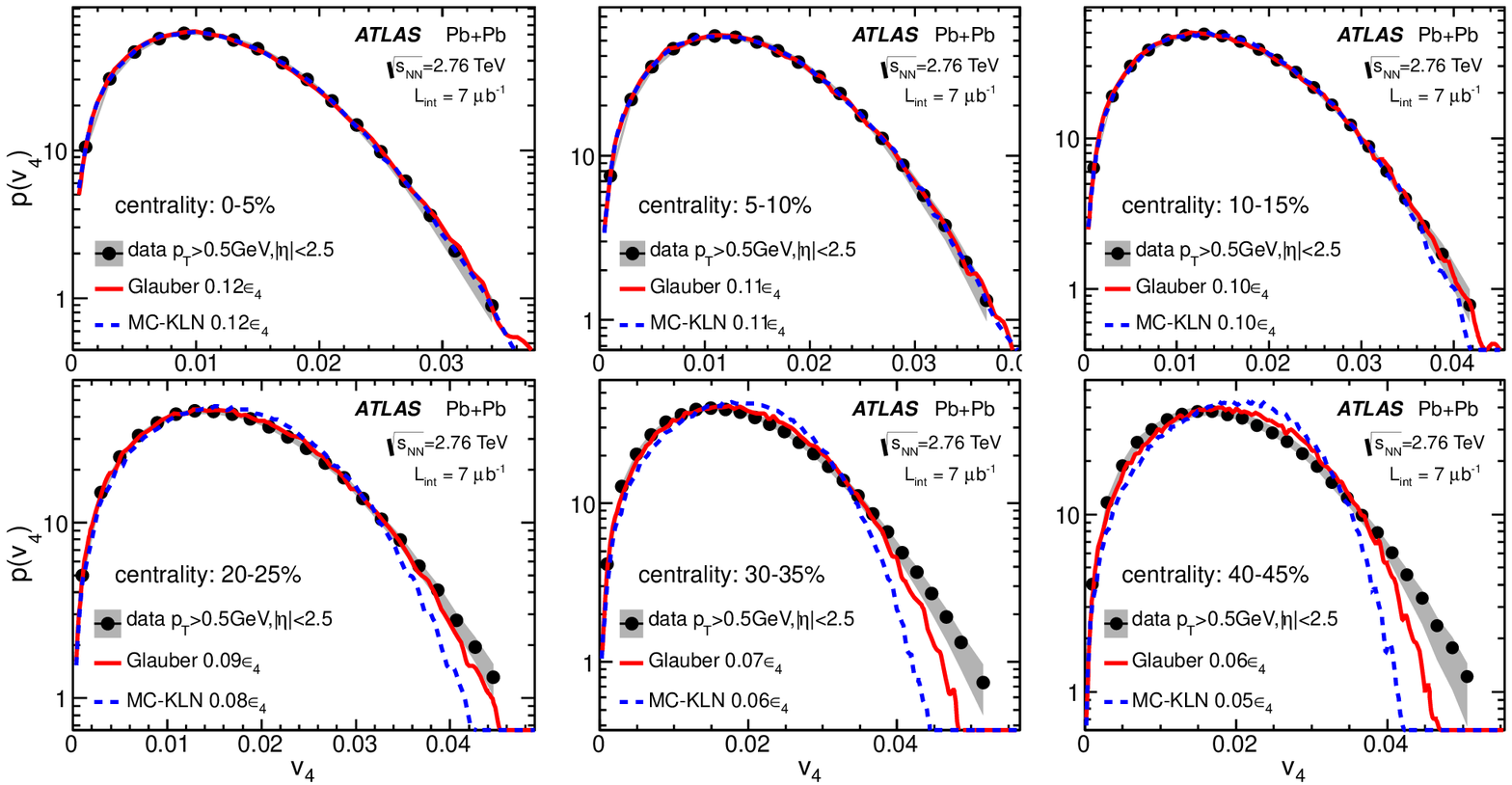}
\caption{\label{fig:result4c} The EbyE $v_4$ distributions compared with the $\epsilon_4$ distributions from two initial geometry models: a Glauber model (solid lines) and the MC-KLN model (dashed lines). The $\epsilon_4$ distributions have been rescaled to the same mean values. The scale factors are indicated in the legends.}
\end{figure}
Figure~\ref{fig:result4a} compares the EbyE $v_2$ distributions with the distributions of the eccentricity $\epsilon_2$ of the initial geometry, calculated via eq.~(\ref{eq:ena}) from the Glauber model~\cite{Miller:2007ri} and the MC-KLN model~\cite{Drescher:2006pi}. The MC-KLN model is based on the Glauber model but takes into account the corrections to the initial geometry due to gluon saturation effects. Three million events have been generated and grouped into centrality intervals according to the impact parameter. The $\epsilon_2$ distribution for each centrality interval is rescaled to match the $\langle v_2\rangle$ of the data, and then normalized to form a probability density function. Since $v_2$ is expected to be proportional to $\epsilon_2$ in most hydrodynamic calculations~\cite{Qiu:2011iv}, the deviations between the $v_2$ distributions and the rescaled $\epsilon_2$ distributions can be used to improve the modeling of the initial geometry. Figure~\ref{fig:result4a} shows that the rescaled $\epsilon_2$ distributions describe the data well for the most central collisions, but not so well for non-central collisions. In peripheral collisions, both the Glauber and MC-KLN models fail to describe the data. A smaller scale factor is generally required for the MC-KLN model, reflecting the fact that the $\epsilon_2$ values from the MC-KLN model are on average larger than those from the Glauber model. Similar comparisons between $v_n$ and $\epsilon_n$ for $n=3$ and $n=4$ are shown in figure~\ref{fig:result4b} and figure~\ref{fig:result4c}, respectively. Agreement with the models is better than in the $n=2$ case. However, this could simply reflect the fact that these distributions are dominated by Gaussian fluctuations, which have a universal shape. The shape differences between the $v_n$ and $\epsilon_n$ distributions are also quantified in the right panels of figure~\ref{fig:result2} by comparing the values of $\sigma_{v_n}/\langle v_n\rangle$ with the values of $\sigma_{\epsilon_n}/\langle \epsilon_n\rangle$. Clearly, both the Glauber and MC-KLN models fail to describe the data consistently, in particular for $n=2$, across most of the measured centrality range. It should be noted that in the centrality intervals that correspond to the more peripheral collisions (e.g centrality interval 55--60\%), the $\epsilon_n$ values have been scaled down by a large factor and the sharp cutoffs of the $\epsilon_n$ distributions are a natural consequence of the kinematic constraint $\epsilon_n\leq1$ (see eq.~(\ref{eq:ena})). However, this behaviour also implies that $v_n$ is not proportional to $\epsilon_n$ for large $\epsilon_n$ values. 

\section{Summary}
\label{sec:con}
Measurements of the event-by-event harmonic flow coefficients $v_n$ for $n=2,3$ and 4 have been performed using 7 $\mu$b$^{-1}$ of Pb+Pb collision data at $\sqrt{s_{_{NN}}}=2.76$ TeV collected by the ATLAS experiment at the LHC. The observed $v_n$ distributions are measured using charged particles in the pseudorapidity range $|\eta|<2.5$ and the transverse momentum range $\pT>0.5$~GeV, which are then unfolded via a response function to estimate the true $v_n$ distributions. The response function is constructed via a data-driven method, which maps the true $v_n$ distribution to the observed $v_n$ distribution. The influence of residual non-flow effects are studied by varying the pseudorapidity gap between the two subevents used to obtain the observed $v_n$ distributions and the response functions, as well as by comparing the two different analysis methods based either on the single particle distribution or two-particle correlations. The influence of the residual non-flow is also compared to an estimation based on model simulations~\cite{Jia:2013tja}, and the latter is found to be within the total systematic uncertainty of this measurement.

The $v_n$ distributions are obtained in various centrality intervals: over the 0-70\% centrality range for $v_2$, 0-60\% centrality range for $v_3$ and 0-45\% centrality range for $v_4$. The measured $v_2$ distributions are found to approach that of a radial projection of a 2D Gaussian distribution centred around zero in the 0--2\% centrality range, which is consistent with a scenario where fluctuations are the primary contribution to the overall shape (fluctuation-only scenario) for these most central collisions. Starting with the centrality interval 5-10\%, the $v_2$ distributions differ significantly from this scenario, suggesting that they have a significant component associated with the average collision geometry in the reaction plane, $v_2^{\mathrm{RP}}$. In contrast, the $v_3$ and $v_4$ distributions are consistent with a pure 2D Gaussian-fluctuation scenario (i.e., $v_n^{\mathrm{RP}}=0$) over most of the measured centrality range. However, a systematic deviation from this fluctuation-only scenario is observed for $v_3$ in mid-central collisions, the presence of a non-zero $v_3^{\mathrm{\;RP}}$ can be allowed. Similarly, due to the limited range of the measured $v_4$ distribution, a non-zero $v_4^{\mathrm{\;RP}}$ on the order of $\delta_{_{v_4}}$ can not be excluded by this analysis.

The $v_n$ distributions are also measured separately for charged particles with $0.5<\pT<1$~GeV and $\pT>1$~GeV. The shape of the unfolded distributions, when rescaled to the same $\langle v_n\rangle$, is found to be nearly the same for the two $\pT$ ranges. This finding suggests that the hydrodynamic response to the eccentricity of the initial geometry has little variation in this $\pT$ region. The ratios of the width to the mean, $\sigma_{v_n}/\langle v_n\rangle$, of these distributions are studied as a function of the average number of participating nucleons ($\langle N_{\mathrm{part}}\rangle$) and $\pT$. The values of $\sigma_{v_2}/\langle v_2\rangle$ are observed to reach a minimum of 0.34 for $\langle N_{\mathrm{part}}\rangle\approx200$, while the values of $\sigma_{v_3}/\langle v_3\rangle$ and $\sigma_{v_4}/\langle v_4\rangle$ are nearly independent of $\langle N_{\mathrm{part}}\rangle$ and are close to the value expected for a pure Gaussian-fluctuation scenario.

To understand further the role of average geometry and fluctuations for $v_2$, the $v_2$ distributions have been fitted to a Bessel--Gaussian function to estimate the value of $v_2^{\mathrm{RP}}$ and the width of the fluctuation $\delta_{_{v_2}}$. In central collisions, where $v_2^{\mathrm{RP}}$ values are small and change rapidly with $\langle N_{\mathrm{part}}\rangle$, narrow binning in centrality is necessary in order to obtain reliable estimates for these parameters. The values of $\delta_{_{v_2}}/v_2^{\mathrm{RP}}$ are found to decrease with $\langle N_{\mathrm{part}}\rangle$ and reach a minimum of $0.38\pm0.02$ at $\left\langle N_{\mathrm{part}}\right\rangle\approx200$, but then increase and are greater than one in central collisions. Furthermore, a systematic deviation of the fit from the data is observed for centralities starting in the 15-20\% centrality interval, and becoming more pronounced for the more peripheral collisions, suggesting significant non-Gaussian behaviour in the flow fluctuations for collisions with small $\langle N_{\mathrm{part}}\rangle$. Multi-particle cumulant methods have been used to study such non-Gaussian behaviour. However, the $v_2$ coefficients for the four-, six- and eight-particle cumulants calculated from the $v_2$ distribution are found to agree with each other over the full centrality range to within 0.5\%-2\%. Hence the precision of experimental measurements of higher-order cumulants needs to be better than a few percent in order to be sensitive to the non-Gaussian behaviour in the $v_n$ distributions.

To elucidate the relation between the azimuthal anisotropy and underlying collision geometry, the measured $v_n$ distributions are compared with the eccentricity distributions of the initial geometry from the Glauber model and the MC-KLN model. Both models fail to describe the data consistently over most of the measured centrality range. 


\section{Acknowledgements}

We thank CERN for the very successful operation of the LHC, as well as the
support staff from our institutions without whom ATLAS could not be
operated efficiently.

We acknowledge the support of ANPCyT, Argentina; YerPhI, Armenia; ARC,
Australia; BMWF and FWF, Austria; ANAS, Azerbaijan; SSTC, Belarus; CNPq and FAPESP,
Brazil; NSERC, NRC and CFI, Canada; CERN; CONICYT, Chile; CAS, MOST and NSFC,
China; COLCIENCIAS, Colombia; MSMT CR, MPO CR and VSC CR, Czech Republic;
DNRF, DNSRC and Lundbeck Foundation, Denmark; EPLANET, ERC and NSRF, European Union;
IN2P3-CNRS, CEA-DSM/IRFU, France; GNSF, Georgia; BMBF, DFG, HGF, MPG and AvH
Foundation, Germany; GSRT and NSRF, Greece; ISF, MINERVA, GIF, DIP and Benoziyo Center,
Israel; INFN, Italy; MEXT and JSPS, Japan; CNRST, Morocco; FOM and NWO,
Netherlands; BRF and RCN, Norway; MNiSW, Poland; GRICES and FCT, Portugal; MERYS
(MECTS), Romania; MES of Russia and ROSATOM, Russian Federation; JINR; MSTD,
Serbia; MSSR, Slovakia; ARRS and MIZ\v{S}, Slovenia; DST/NRF, South Africa;
MICINN, Spain; SRC and Wallenberg Foundation, Sweden; SER, SNSF and Cantons of
Bern and Geneva, Switzerland; NSC, Taiwan; TAEK, Turkey; STFC, the Royal
Society and Leverhulme Trust, United Kingdom; DOE and NSF, United States of
America.

The crucial computing support from all WLCG partners is acknowledged
gratefully, in particular from CERN and the ATLAS Tier-1 facilities at
TRIUMF (Canada), NDGF (Denmark, Norway, Sweden), CC-IN2P3 (France),
KIT/GridKA (Germany), INFN-CNAF (Italy), NL-T1 (Netherlands), PIC (Spain),
ASGC (Taiwan), RAL (UK) and BNL (USA) and in the Tier-2 facilities
worldwide.
\bibliographystyle{JHEP}

\providecommand{\href}[2]{#2}\begingroup\raggedright\endgroup

\appendix

\section{Comprehensive performance and data plots}
\begin{figure}[!h]
\centering
\includegraphics[width=1\columnwidth]{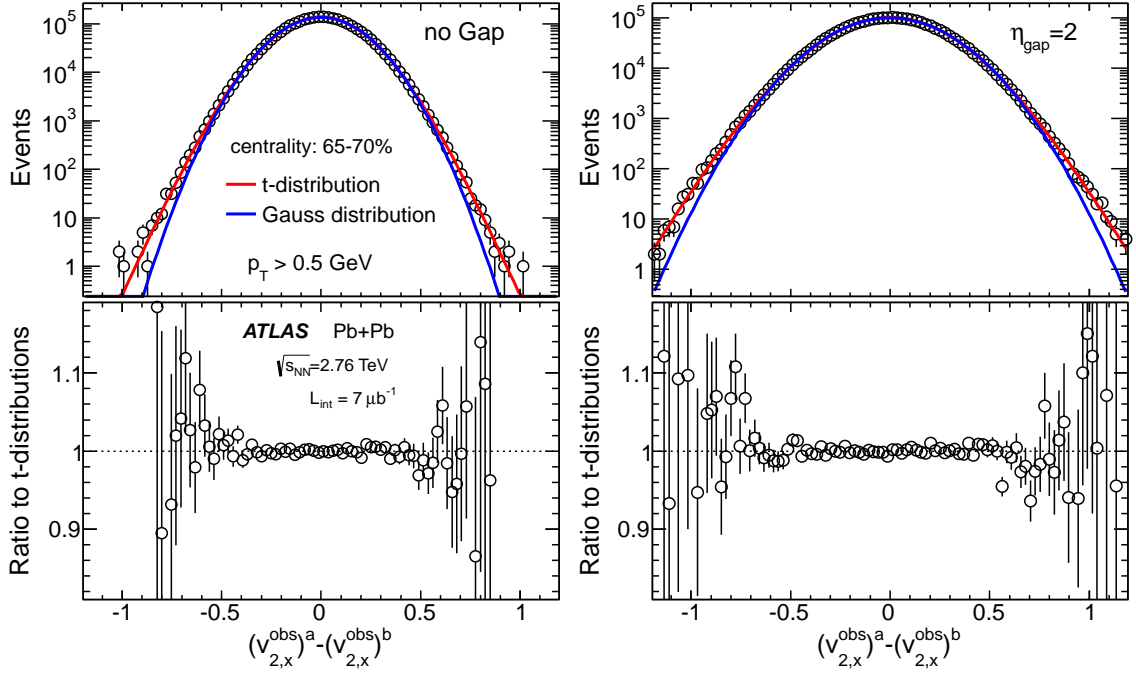}
\caption{\label{fig:rp} Top panels: the $x$-projection of the differences of $\protect\overrightharp{v}_2$ vectors calculated for the two half-IDs $p_{\mathrm{sub}}\left((\protect\overrightharp{v}_2^{\;\mathrm{obs}})^{\rm a}-(\protect\overrightharp{v}_2^{\;\mathrm{obs}})^{\rm b}\right)$ (arbitrary normalization) for the 65--70\% centrality interval with $\eta_{\mathrm{gap}}=0$ (left panel) and $\eta_{\mathrm{gap}}=2$ (right panel), together with fits to a Gaussian function and a Student's t-distribution. Bottom panels: the ratios of the data to the t-distribution fits. The distributions are wider for $\eta_{\mathrm{gap}}=2$ since the average number of tracks used is smaller.}
\end{figure}

\begin{figure}[!h]
\centering
\includegraphics[width=1\columnwidth]{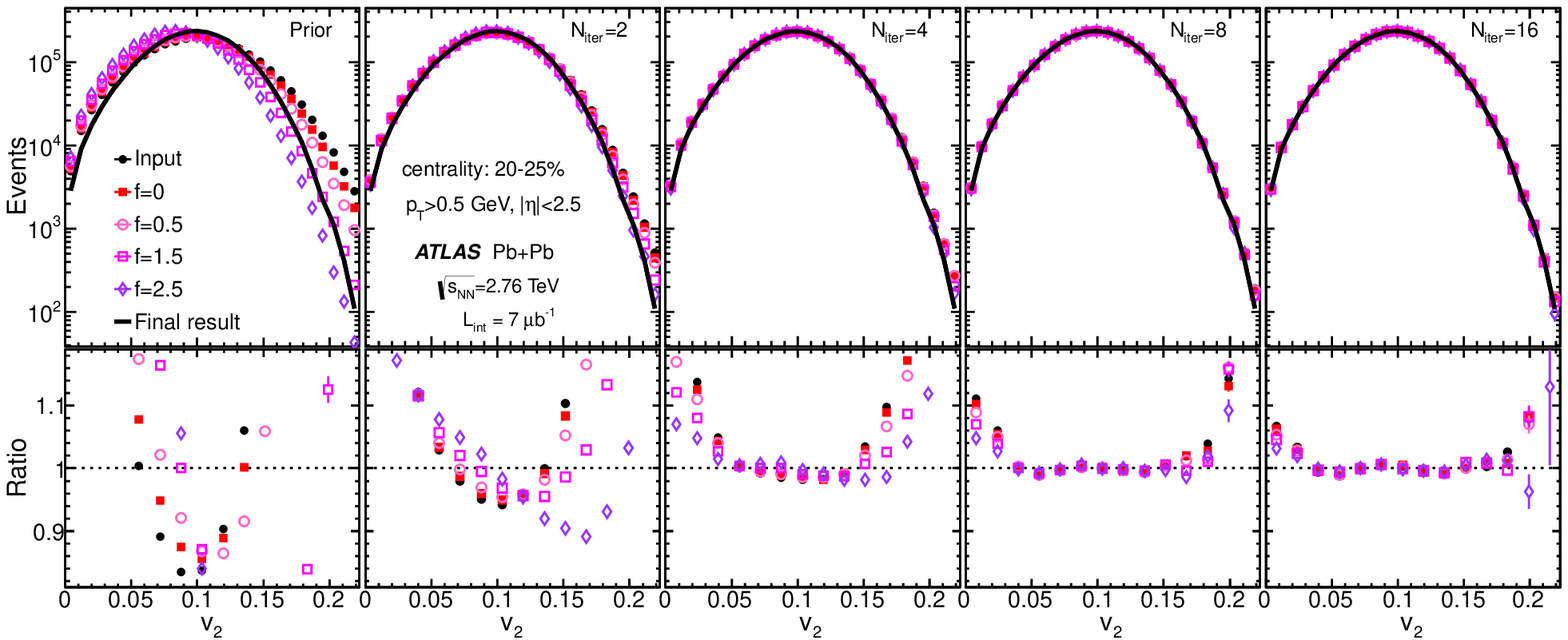}
\caption{\label{fig:perf4a}  Convergence behaviour of $v_2$ in the 20--25\% centrality interval for five choices of priors for several values of $N_{\mathrm{iter}}$, increasing from left to right. The top panels show the distributions after a certain number of iterations and bottom panels show the ratios to the result for $N_{\mathrm{iter}}=128$. A common reference, shown by the solid lines in the top panels, is calculated by averaging the result for $f=0$ and $f=0.5$ (eq.~(\ref{eq:pri0})) with $N_{\mathrm{iter}}=128$. }
\end{figure}

\begin{figure}[!h]
\centering
\includegraphics[width=1\columnwidth]{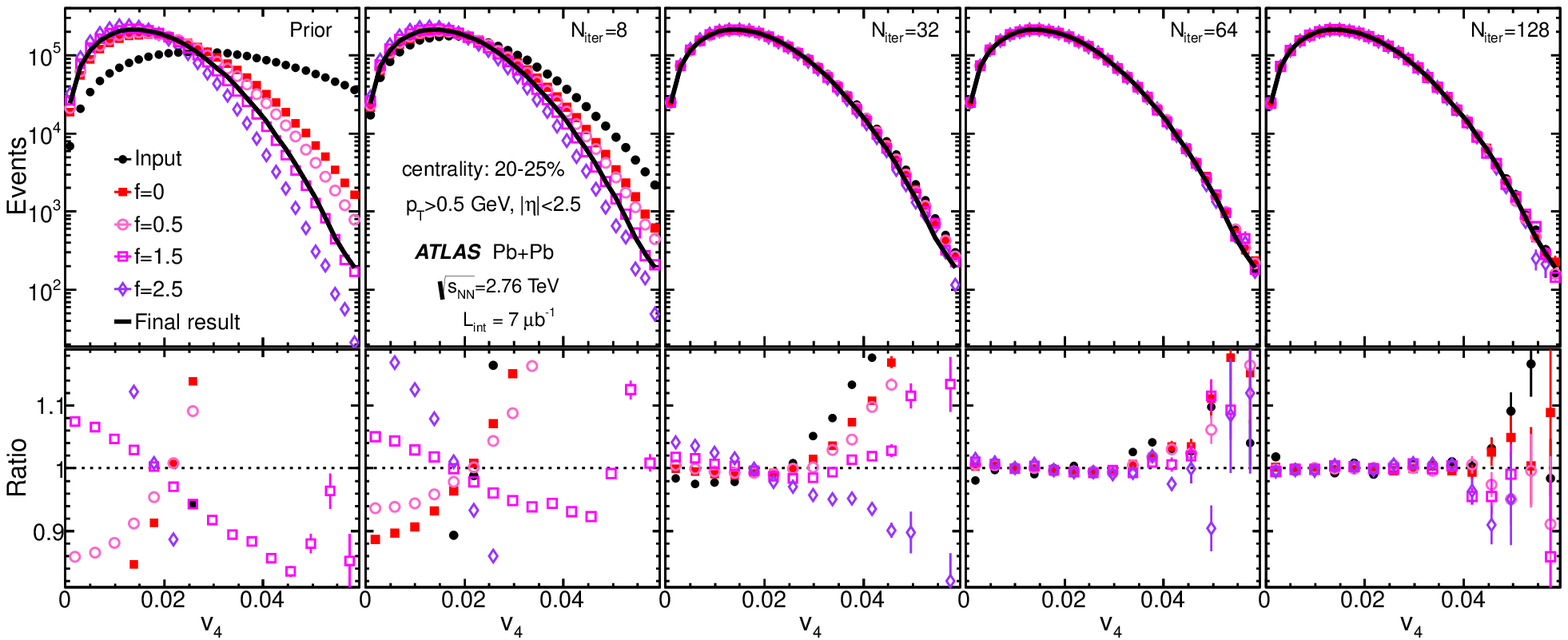}
\caption{\label{fig:perf4b}  Convergence behaviour of $v_4$ in the 20--25\% centrality interval for five choices of priors for several values of $N_{\mathrm{iter}}$, increasing from left to right. The top panels show the distributions after a certain number of iterations and bottom panels show the ratios to the result for $N_{\mathrm{iter}}=128$. A common reference, shown by the solid lines in the top panels, is calculated by averaging the result for $f=0$ and $f=0.5$ (eq.~(\ref{eq:pri0})) with $N_{\mathrm{iter}}=128$. }
\end{figure}

\begin{figure}[!h]
\centering
\includegraphics[width=1\columnwidth]{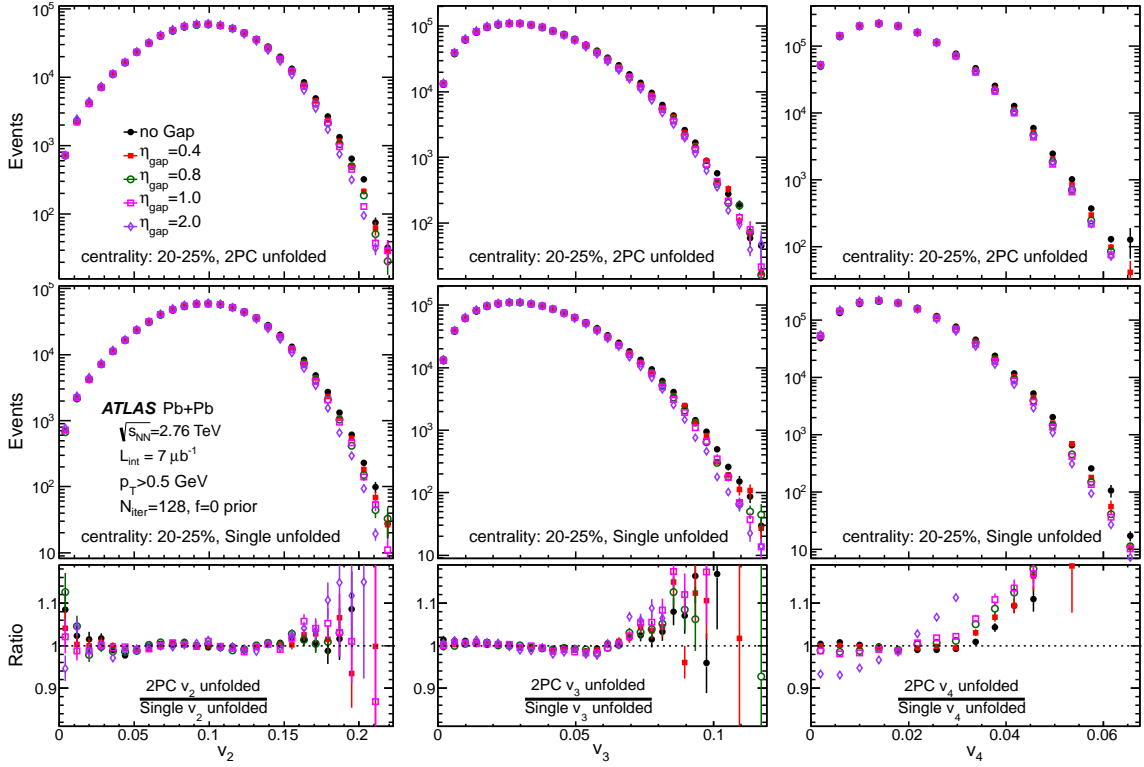}
\caption{\label{fig:perf6a}  The unfolded distributions from 2PC method (top row), single-particle method (middle row) and the ratios of the two (bottom row) for different values of $\eta_{\mathrm{gap}}$ in the 20--25\% centrality interval and for $n=2$ (left panels), $n=3$ (middle panels) and $n=4$ (right panels). }
\end{figure}

\begin{figure}[!h]
\centering
\includegraphics[width=1\columnwidth]{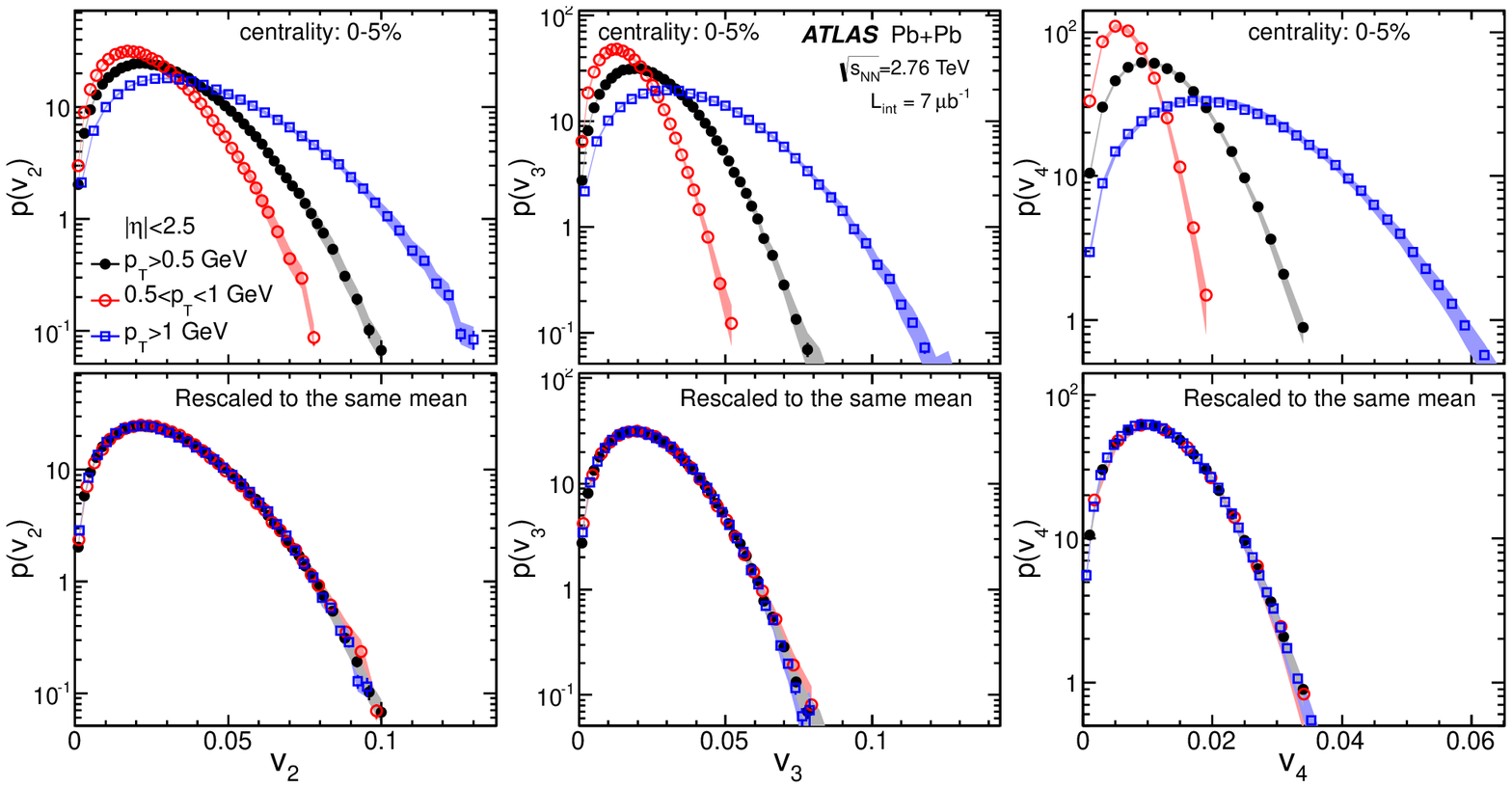}
\caption{\label{fig:perf7a}  Top panels: The unfolded distributions for $v_n$ in the 0--5\% centrality interval for charged particles in the $\pT>0.5$~GeV, $0.5<\pT<1$~GeV and $\pT>1$~GeV ranges. Bottom panels: same distributions but rescaled horizontally so the $\langle v_n\rangle$ values match that for the $\pT>0.5$~GeV range.  The shaded bands represent the systematic uncertainties on the $v_n$-shape.}
\end{figure}

\begin{figure}[!h]
\centering
\includegraphics[width=1\columnwidth]{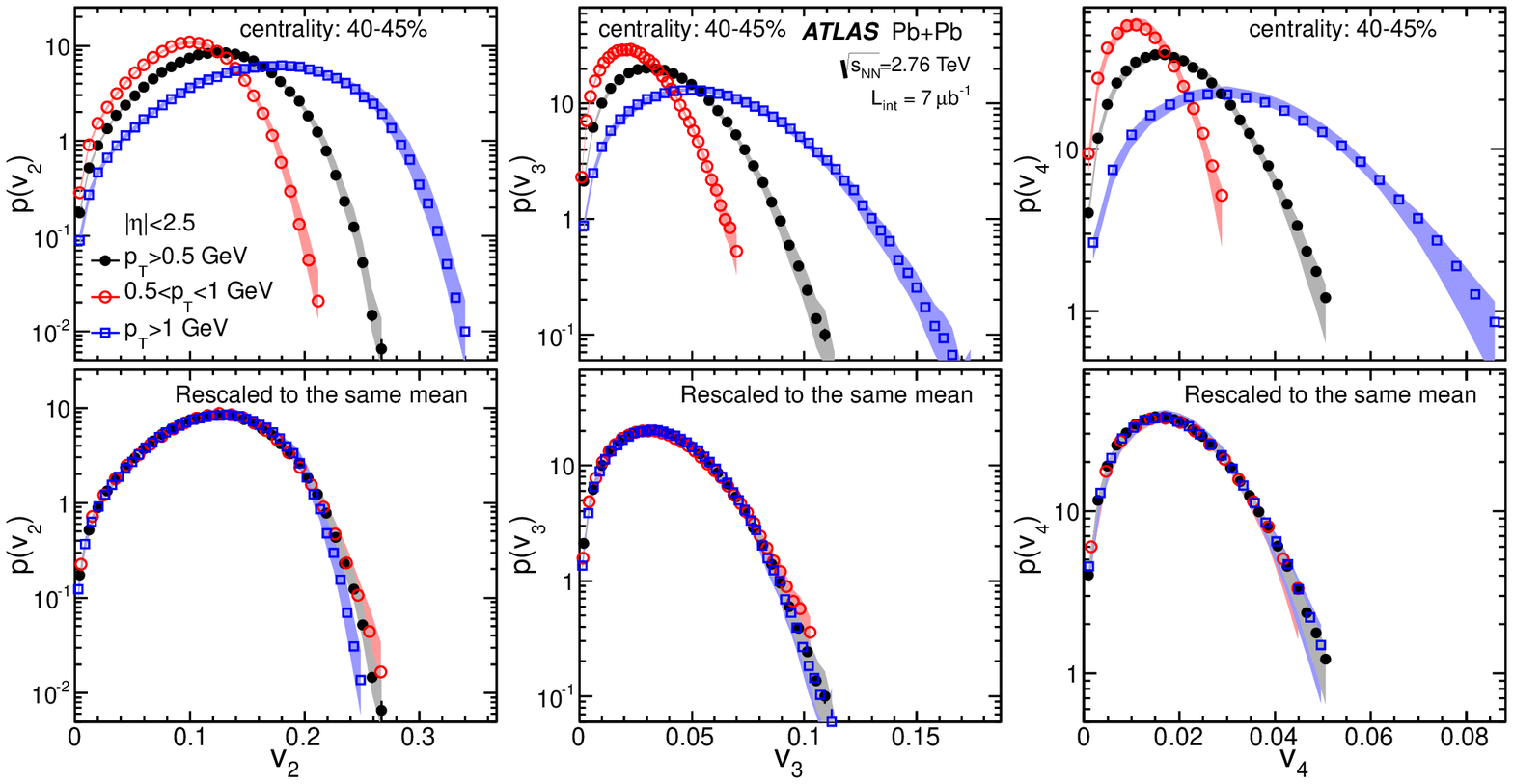}
\caption{\label{fig:perf7b}  Top panels: The unfolded distributions for $v_n$ in the 40--45\% centrality interval for charged particles in the $\pT>0.5$~GeV, $0.5<\pT<1$~GeV and $\pT>1$~GeV ranges. Bottom panels: same distributions but rescaled horizontally so the $\langle v_n\rangle$ values match that for the $\pT>0.5$~GeV range.  The shaded bands represent the systematic uncertainties on the $v_n$-shape.}
\end{figure}

\begin{figure}[!h]
\centering
\includegraphics[width=1\columnwidth]{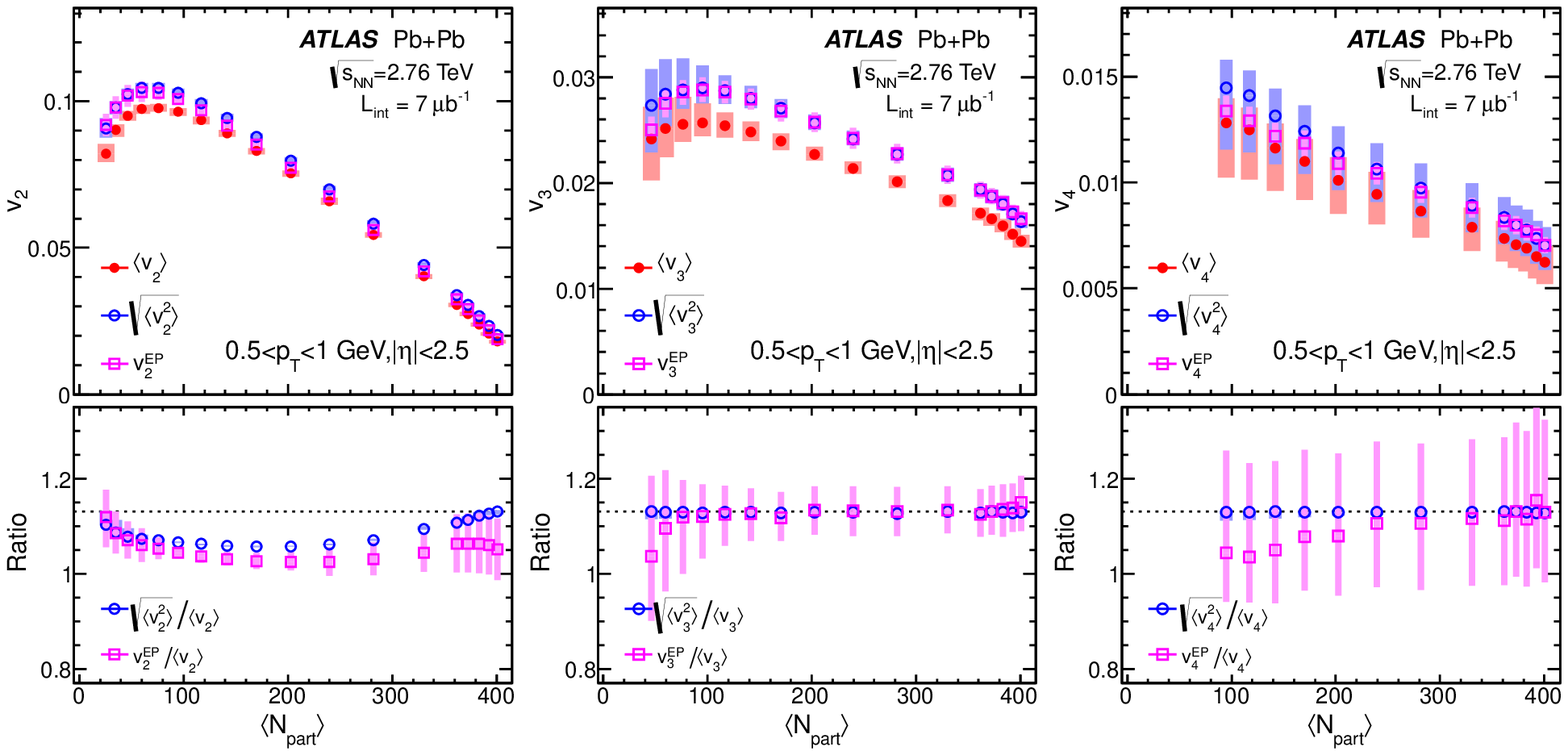}
\caption{\label{fig:result3b} Top panels: Comparison of $\langle v_n\rangle$ and $\sqrt{\langle v_n^2\rangle}\equiv\sqrt{\langle v_n\rangle^2+\sigma_{v_n}^2}$, derived from the EbyE $v_n$ distributions, with the $v_n^{\mathrm{EP}}$~\cite{Aad:2012bu}, for charged particles in the $0.5<\pT<1$~GeV range. Bottom panels: the ratios of $\sqrt{\langle v_n^2\rangle}$ and  $v_n^{\mathrm{EP}}$ to $\langle v_n\rangle$. The shaded bands represent the systematic uncertainties. The dotted lines in bottom panels indicate $\sqrt{\langle v_n^2\rangle}/\langle v_n\rangle=1.13$, the value expected for the radial projection of a 2D Gaussian distribution (eq.~(\ref{eq:large})).}
\end{figure}

\begin{figure}[!h]
\centering
\includegraphics[width=1\columnwidth]{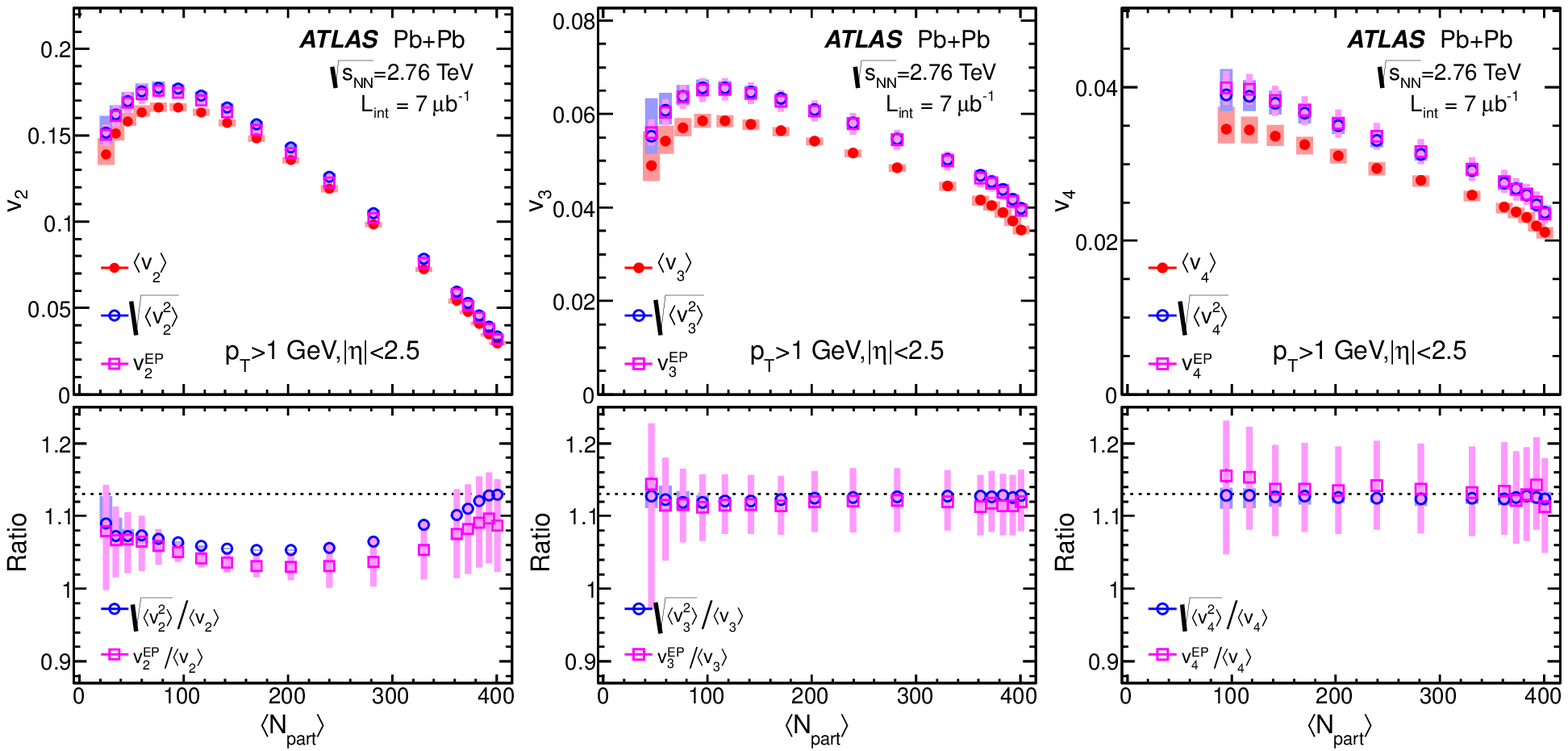}
\caption{\label{fig:result3c} Top panels: Comparison of $\langle v_n\rangle$ and $\sqrt{\langle v_n^2\rangle}\equiv\sqrt{\langle v_n\rangle^2+\sigma_{v_n}^2}$, derived from the EbyE $v_n$ distributions, with the $v_n^{\mathrm{EP}}$~\cite{Aad:2012bu}, for charged particles in the $\pT>1$~GeV range. Bottom panels: the ratios of $\sqrt{\langle v_n^2\rangle}$ and $v_n^{\mathrm{EP}}$ to $\langle v_n\rangle$. The shaded bands represent the systematic uncertainties. The dotted lines in bottom panels indicate $\sqrt{\langle v_n^2\rangle}/\langle v_n\rangle=1.13$, the value expected for the radial projection of a 2D Gaussian distribution (eq.~(\ref{eq:large})).}
\end{figure}

For reference, a more complete set of plots detailing the unfolding performance and cross-checks is presented. Figure~\ref{fig:rp} shows the distribution of the difference of the flow vectors $\protect\overrightharp{v}_2^{\;\mathrm{obs}}$ obtained for two half-IDs in peripheral collisions, calculated either without an $\eta$ gap or with $\eta_{\mathrm{gap}}=2$ between the two half-IDs. As mentioned in section~\ref{sec:m1.1},  due to the small number of charged particles in the events, this distribution is expected to be described by the Student's t-distribution instead of the Gaussian distribution. Figures~\ref{fig:perf4a} and \ref{fig:perf4b} show the dependence on the priors for $v_2$ and $v_4$ in the 20--25\% centrality interval. Despite the very different shapes of the underlying flow distributions, the convergence is robust and independent of the prior. Figure~\ref{fig:perf6a} compares the unfolding performance for the single-particle and 2PC methods for several choices of $\eta_{\mathrm{gap}}$ and demonstrates the consistency of the two methods for all $\eta_{\mathrm{gap}}$ values considered. Figures~\ref{fig:perf7a} and \ref{fig:perf7b} are similar to figure~\ref{fig:perf7} but are obtained for other centrality intervals. They demonstrate that the shapes of the $v_n$ distributions are the same for $0.5<\pT<1$~GeV and $\pT>1$~GeV ranges in all centrality intervals. Figures~\ref{fig:result3b} and \ref{fig:result3c} show a comparison between $\langle v_n\rangle$, $\sqrt{\langle v_n^2\rangle}$ and $v_{n}^{\mathrm{EP}}$ in two different $\pT$ ranges; the trends are similar to those seen in figure~\ref{fig:result3a}.

\begin{figure}[!h]
\centering
\includegraphics[width=0.5\columnwidth]{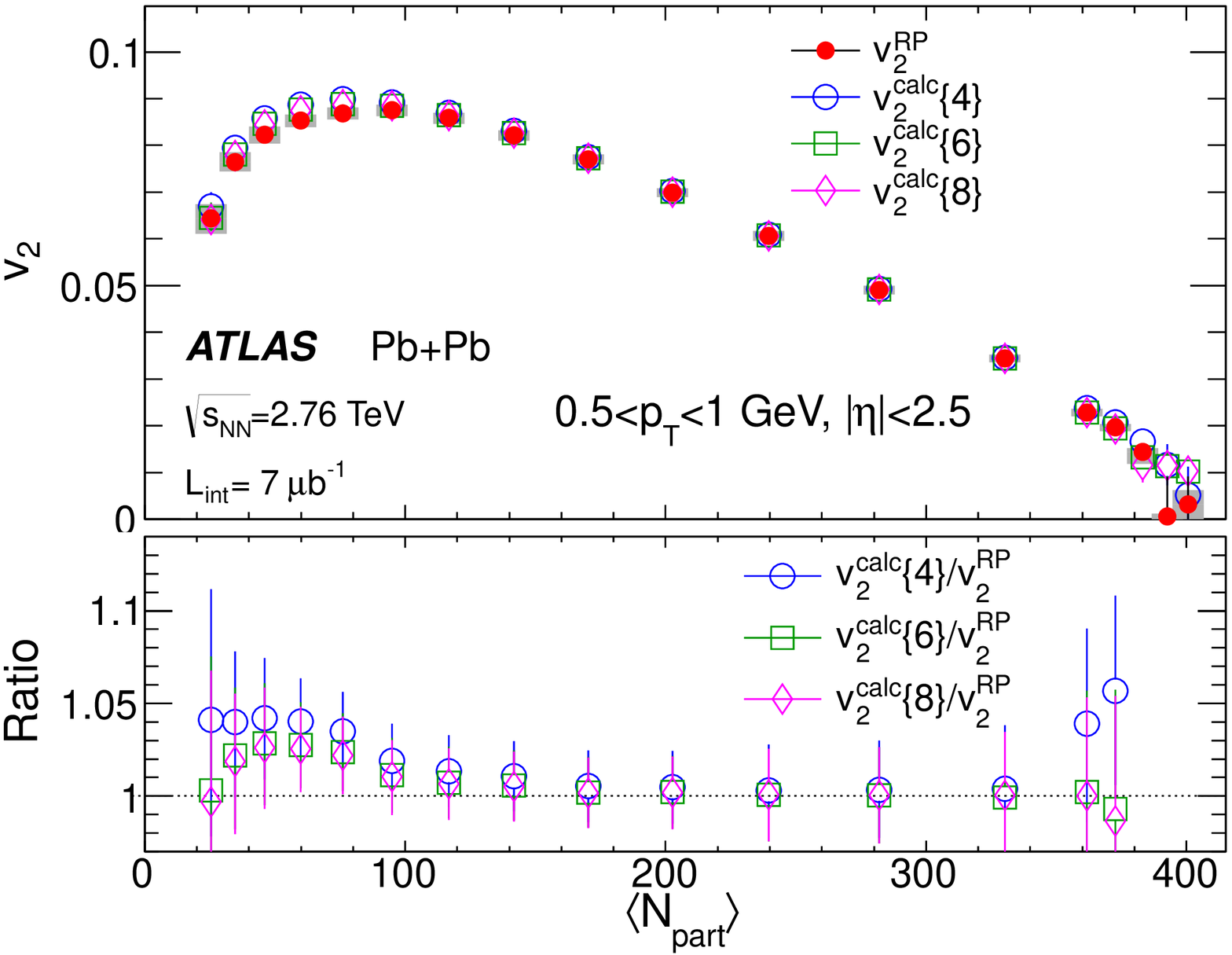}\includegraphics[width=0.5\columnwidth]{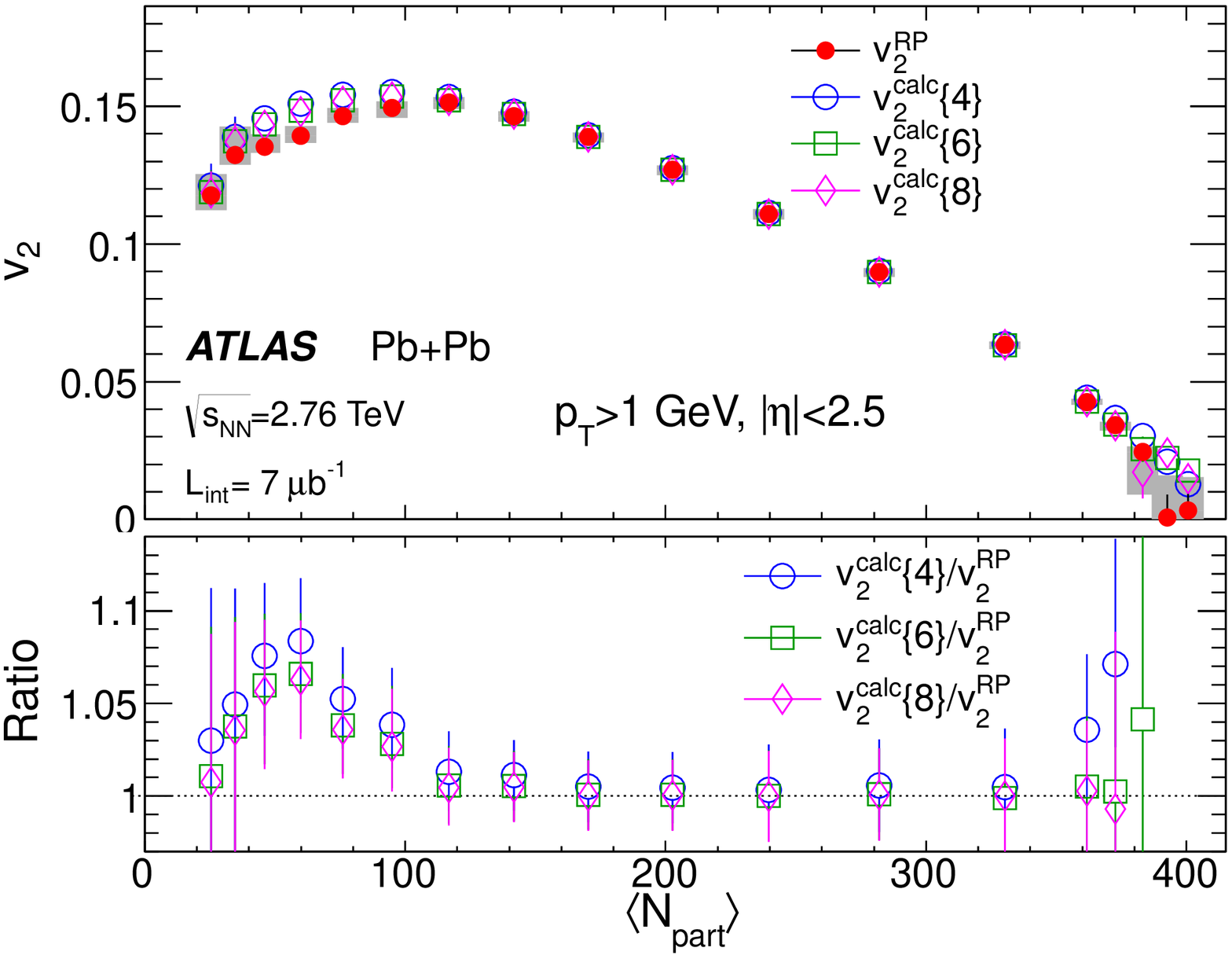}
\caption{\label{fig:fit3all} Comparison in $0.5<\pT<1$ GeV (left panel) and $\pT>1$ GeV (right panel) of the $v_2^{\mathrm{RP}}$ obtained from the Bessel--Gaussian fit of the $v_2$ distributions with the values for four-particle ($v_2^{\mathrm{calc}}\{4\}$), six-particle ($v_2^{\mathrm{calc}}\{6\}$) and eight-particle ($v_2^{\mathrm{calc}}\{8\}$) cumulants calculated directly from the $v_2$ distributions via eq.~(\ref{eq:cumu2}). The bottom part of each panel shows the ratios of the cumulants to the fit results, with the error bars representing the total uncertainties.}
\end{figure}

\begin{figure}[!h]
\centering
\includegraphics[width=1\columnwidth]{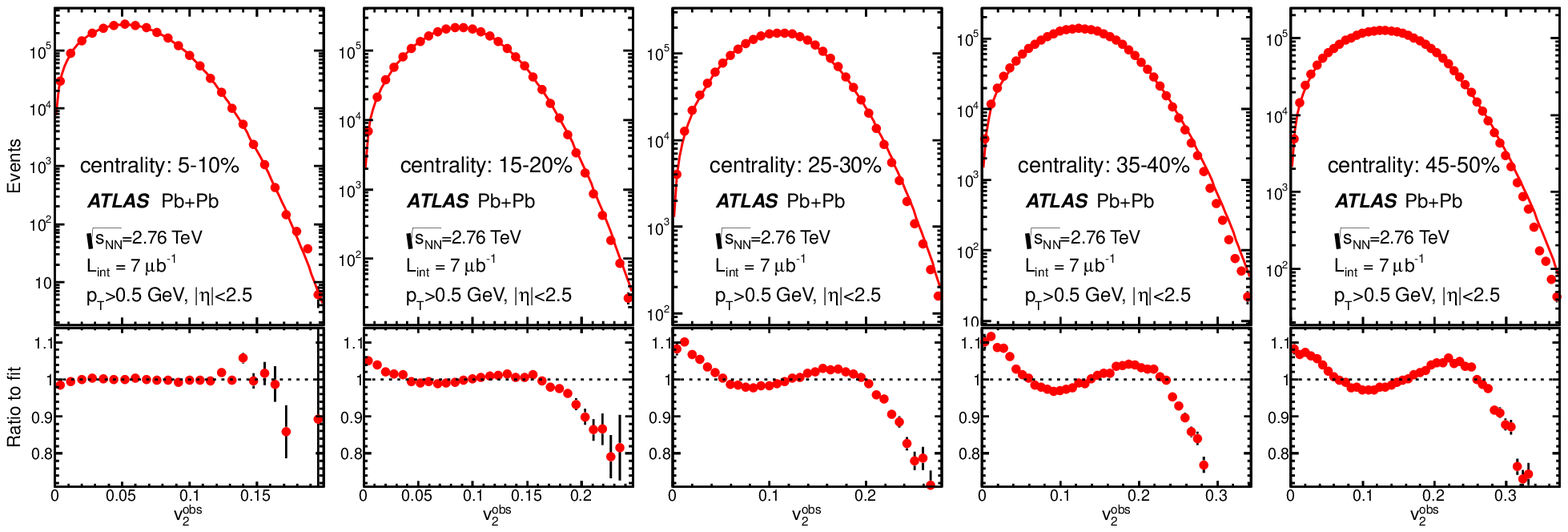}
\caption{\label{fig:fitbefore} The $v_2^{\mathrm{obs}}$ distributions before unfolding (top) and their ratios to the Bessel--Gaussian fits (bottom) in various centrality ranges (see legends).}
\end{figure}

\begin{figure}[!h]
\centering
\includegraphics[width=1\columnwidth]{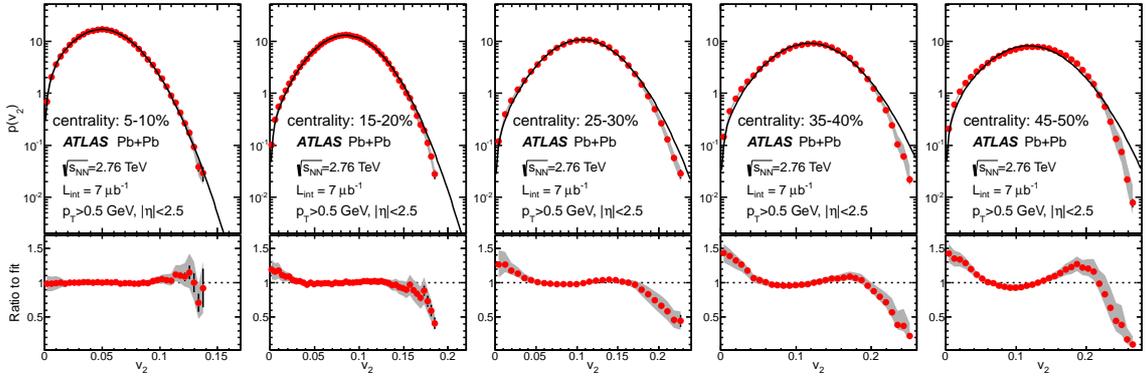}
\caption{\label{fig:fitafter} The final $v_2$ distributions obtained from unfolding (top) and their ratios to the Bessel--Gaussian fits (bottom) in various centrality ranges (see legends).}
\end{figure}

\begin{figure}[!h]
\centering
\includegraphics[width=0.8\columnwidth]{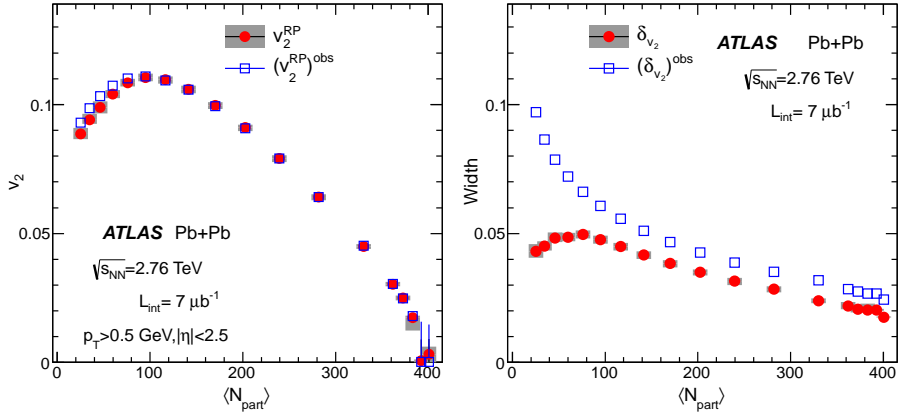}
\caption{\label{fig:fitpar} The values of $v_2^{\mathrm{RP}}$ and $(v_2^{\mathrm{RP}})^{\mathrm{obs}}$ (left panel) and the values of $\delta_{v_2}$ and $(\delta_{v_2})^{\mathrm{obs}}$ (right panel) obtained from the Bessel-Gaussian fits to the $v_2$ and $v_2^{\mathrm{obs}}$ distributions.}
\end{figure}

Figures~\ref{fig:fit3all} compares the $v_2^{\mathrm{RP}}$ from the Bessel--Gaussian fit of the $v_2$ distribution with $v_2^{\mathrm{calc}}\{2k\}$ values for $0.5<\pT<1$~GeV and $\pT>1$~GeV ranges. The trends are similar to those seen in figure~\ref{fig:fit3}. However a slightly bigger deviation between $v_2^{\mathrm{RP}}$ and $v_2^{\mathrm{calc}}\{2k\}$ is observed in peripheral collisions for the $\pT>1$~GeV range. Figures~\ref{fig:fitbefore} and ~\ref{fig:fitafter} show the Bessel-Gaussian fits to the $v_2^{\;\mathrm{obs}}$ and $v_2$ distributions in several centrality intervals. Deviations from the fits are observed in both types of distributions with a similar trend, but the deviations are smaller for the $v_2^{\;\mathrm{obs}}$ distributions. Figure~\ref{fig:fitpar} compares the centrality dependence of $v_2^{\mathrm{RP}}$ and $\delta_{v_2}$ from the fits to distributions before and after unfolding. The effects of smearing by the response function increase the values for $\delta_{v_2}$, but the values of $v_2^{\mathrm{RP}}$ only increase slightly. This is expected since the true $v_2$ distributions in the bulk region are close to a Bessel-Gaussian shape and the smearing due to the response function is nearly Gaussian (see eqs.~\ref{eq:fluc2} and \ref{eq:5b}). The slightly larger $v_2^{\mathrm{RP}}$ values for the $v_2^{\;\mathrm{obs}}$ distribution may reflect non-flow contributions in $v_2^{\;\mathrm{obs}}$ or deviation of the response function from a Gaussian (figure~~\ref{fig:rp}).

\begin{figure}[!h]
\centering
\includegraphics[width=0.9\columnwidth]{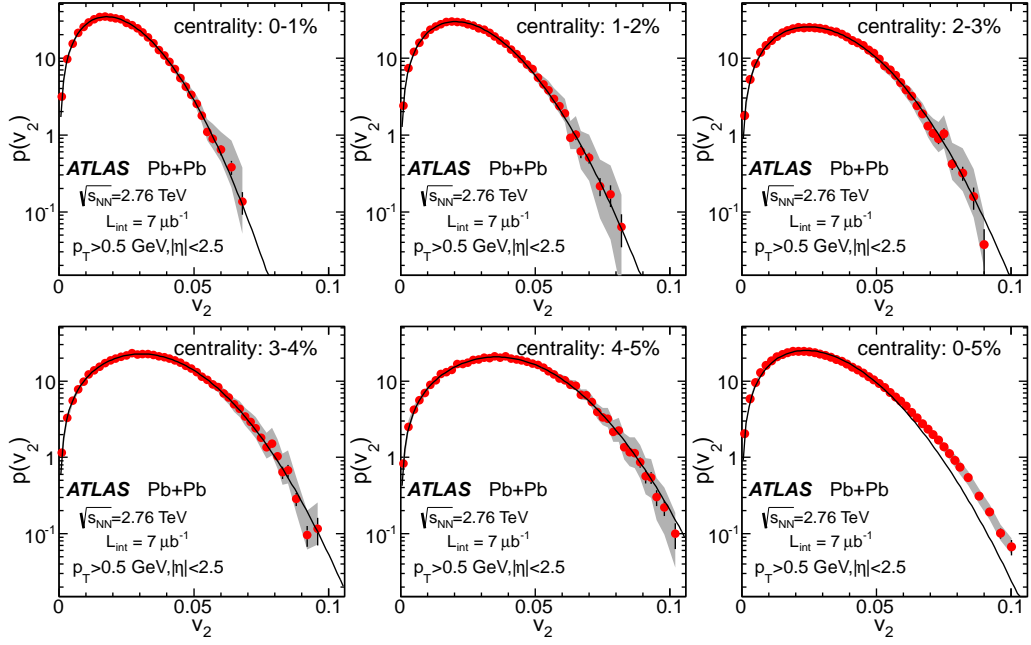}
\caption{\label{fig:pp3} The probability density distributions of $v_2$ for the 0--5\% centrality interval and the five individual 1\% centrality ranges, together with the fit to the Bessel--Gaussian function.}
\end{figure}

Figure~\ref{fig:pp3} shows the probability density distributions of $v_2$ for the full 0--5\% centrality interval and the five individual 1\% centrality intervals, together with the fits to the Bessel--Gaussian function. The deviation of the data from the Bessel--Gaussian description for the 0-5\% centrality interval is due mainly to the rapid change of $v_2^{\mathrm{RP}}$ with centrality within this interval.  This supports the need to use small centrality intervals for the most central collisions when calculating quantities such as $\langle v_2\rangle$, $\sigma_{v_2}$, $v_2^{\mathrm{RP}}$ and $\delta_{_{v_2}}$ (see figure~\ref{fig:fit2}).




\clearpage 
\begin{flushleft}
{\Large The ATLAS Collaboration}

\bigskip

G.~Aad$^{\rm 48}$,
T.~Abajyan$^{\rm 21}$,
B.~Abbott$^{\rm 112}$,
J.~Abdallah$^{\rm 12}$,
S.~Abdel~Khalek$^{\rm 116}$,
A.A.~Abdelalim$^{\rm 49}$,
O.~Abdinov$^{\rm 11}$,
R.~Aben$^{\rm 106}$,
B.~Abi$^{\rm 113}$,
M.~Abolins$^{\rm 89}$
% ...
\bigskip
\\
% ...
$^{11}$ Institute of Physics, Azerbaijan Academy of Sciences, Baku, Azerbaijan\\
$^{12}$ Institut de F{\'\i}sica d'Altes Energies and Departament de F{\'\i}sica de la Universitat Aut{\`o}noma de Barcelona and ICREA, Barcelona, Spain\\
% ...
$^{21}$ Physikalisches Institut, University of Bonn, Bonn, Germany\\
% ...
$^{48}$ Fakult{\"a}t f{\"u}r Mathematik und Physik, Albert-Ludwigs-Universit{\"a}t, Freiburg, Germany\\
$^{49}$ Section de Physique, Universit{\'e} de Gen{\`e}ve, Geneva, Switzerland\\
% ...
$^{89}$ Department of Physics and Astronomy, Michigan State University, East Lansing MI, United States of America\\
% ...
$^{106}$ Nikhef National Institute for Subatomic Physics and University of Amsterdam, Amsterdam, Netherlands\\
% ...
$^{112}$ Homer L. Dodge Department of Physics and Astronomy, University of Oklahoma, Norman OK, United States of America\\
$^{113}$ Department of Physics, Oklahoma State University, Stillwater OK, United States of America\\
% ...
$^{116}$ LAL, Universit{\'e} Paris-Sud and CNRS/IN2P3, Orsay, France\\
% ...
\end{flushleft}

\end{document}